\documentclass[pra,10pt,twocolumn,notitlepage,aps,floatfix,superscriptaddress]{revtex4-1}
\usepackage[utf8]{inputenc}
\usepackage{amsmath,amssymb,bm,comment,graphicx,color}
\usepackage[breaklinks=true,colorlinks=true,
    linkcolor=blue,urlcolor=blue,citecolor=blue]{hyperref}
\usepackage{enumerate}
\usepackage{algpseudocode}
\usepackage{algorithm}
\usepackage{amsmath}
\usepackage{ulem}

\newcommand{\W}[0]{{\mathcal{W}}}
\newcommand{\D}[0]{{\mathbf{D}}}
\newcommand{\bphi}[0]{{\boldsymbol{\phi}}}
\newcommand{\bpsi}[0]{{\boldsymbol{\psi}}}
\newcommand{\tD}[0]{{\boldsymbol{\mathcal{D}}}}

\newcommand{\w}[0]{{\mathbf{w}}}
\newcommand{\X}[0]{{\mathbf{X}}}
\newcommand{\C}[0]{{{\bf c}}}
\newcommand{\mom}[0]{{\bf P}}

\newcommand{\T}[0]{{\bm\Theta}}

\newcommand{\ScoreD}[0]{\nabla\mathcal{S}_\T(\tD|\alpha)}

\newcommand{\bmu}[0]{{\boldsymbol{\mu}}}
\newcommand{\bSig}[0]{{\boldsymbol{\Sigma}}}

\definecolor{asparagus}{rgb}{0.53, 0.66, 0.42}

\definecolor{watermelon}{rgb}{0.89, 0.45, 0.51}

\begin{document}
\title{Agnostic calculation of atomic free energies with the descriptor density of states}
\author{Thomas D Swinburne}
\email{thomas.swinburne@cnrs.fr}
\affiliation{Aix-Marseille Universit\'{e}, CNRS, CINaM UMR 7325, Campus de Luminy, 13288 Marseille, France}
\author{Clovis Lapointe}
\affiliation{Universit\'{e} Paris-Saclay, CEA, Service de recherche en Corrosion et Comportement des Mat\'{e}riaux, SRMP, 91191, Gif-sur-Yvette, France.}
\author{Mihai-Cosmin Marinica}
\email{mihai-cosmin.marinica@cea.fr}
\affiliation{Universit\'{e} Paris-Saclay, CEA, Service de recherche en Corrosion et Comportement des Mat\'{e}riaux, SRMP, 91191, Gif-sur-Yvette, France.}
\date{\today}
\begin{abstract}
    We present a new method to evaluate vibrational free energies of atomic systems without 
    \textit{a priori} specification of an interatomic potential.
    Our model-agnostic approaches leverages descriptors, high-dimensional feature vectors of atomic structure. 
    The entropy of a high-dimensional density, the descriptor density of states, is accurately estimated with conditional score matching. 
    Casting interatomic potentials into a form extensive in descriptor features, we show free energies emerge as the Legendre–Fenchel conjugate of the descriptor entropy, avoiding all high-dimensional integration. 
    The score matching campaign requires less resources than fixed-model sampling and is highly parallel, reducing wall time to a few minutes, with tensor compression schemes allowing lightweight storage. 
    Our model-agnostic estimator returns differentiable free energy predictions over a broad range of potential parameters in microseconds of CPU effort,
    allowing rapid forward and back propagation of potential variations through finite temperature simulations, long-desired for uncertainty quantification and inverse design. 
    We test predictions against thermodynamic integration calculations over a broad range of models for BCC, FCC and A15 phases of W, Mo and Fe at high homologous 
    temperatures. Predictions pass the stringent accuracy threshold of 1-2 meV/atom (1/40-1/20 kcal/mol) for phase prediction with propagated score uncertainties robustly bounding errors. We also demonstrate targeted fine-tuning, reducing the $\alpha-\gamma$ transition temperature in a non-magnetic machine learning model of Fe from 2030 K to 1063 K through back-propagation,
     with no additional sampling. Applications to liquids and fine-tuning foundational models are discussed along 
     with the many problems in computational science which estimate high dimensional integrals.
\end{abstract}

\maketitle
Determination of finite temperature material properties, such as 
phase stability, heat capacity, elastic constants or thermal expansion coefficients is a central goal of condensed matter physics and materials science.  Theoretical predictions are actively sought as experiments are often time-consuming, expensive, and potentially unfeasible at high temperature or pressure. Atomic simulations employing \textit{ab initio} or empirical energy models allow, in principle, purely \textit{in silico} prediction of finite temperature properties. A generic task is to calculate the \textit{free energy} $\mathcal{F}$ of some set of crystalline phases over a range of temperatures and volumes. If low-temperature quantum statistics are neglected, $\mathcal{F}$ can be defined as the logarithm of an integral in thousands of dimensions, the classical partition function $Z$\cite{reif2009fundamentals}. Accurate phase prediction requires tightly converged estimates 
of $\mathcal{F}$, to within 1-2 meV/atom, or 1/40-1/20 kcal/mol. 

Estimating $\mathcal{F}$ thus requires high dimensional integration, one of the most challenging 
tasks in computational science, the central difficulty in e.g. evidence 
calculation\cite{von2011bayesian,bhat2010derivation,lotfi2022bayesian} in Bayesian statistics 
or density estimation\cite{hyvarinen2005estimation} in machine learning. 

A range of specialized techniques to estimate $\mathcal{F}$ 
have been designed over the last few decades, all some form of stratified sampling\cite{rousset2010free} 
from an analytically tractable reference model\cite{zhu2017efficient,grabowski2019ab,Zhong2023,zhu2024accelerating,menon2024electrons}.
While chemical accuracy in free energy estimation traditionally required expensive \textit{ab initio} calculations in a multistep stratified sampling scheme\cite{grabowski2019ab}, modern machine learning interatomic potentials (MLIPs) are becoming 
a viable replacement.
Recent studies\cite{Zhong2023,menon2024electrons,castellano2024machinelearningassistedcanonical} have shown MLIPs can provide near-\textit{ab initio} accurate free energy predictions, especially when fine-tuned for specific phases\cite{grabowski2019ab}.%, albeit for $\mathcal{O}(10-100)$ cost over traditional potentials. 
In the most popular models\cite{Shapeev_MTP,Thompson_snap_2015,lysogorskiy2021performant,goryaeva2021,Nguyen2023}, including recent message-passing neural networks\cite{batatia2023foundation,perez2025uncertaintyquantificationmisspecifiedmachine}, local atomic configurations 
are encoded using (possibly learned) \textit{descriptor} functions that respect physical symmetries of permutation, translation and rotation. Multiple recent works have noted that descriptors are an ideal latent space for generative models of dynamics\cite{swinburne2023prl} or thermodynamic samples, using e.g. normalizing flows\cite{tamagnone2024coarse,ahmad2022free,wirnsberger2022normalizing,noe2019boltzmann} or variational autoencoders\cite{baima2022capabilities} to accelerate the convergence of any free energy estimate.
\begin{figure*}[!t]
    \centering
    \includegraphics[width=\linewidth]{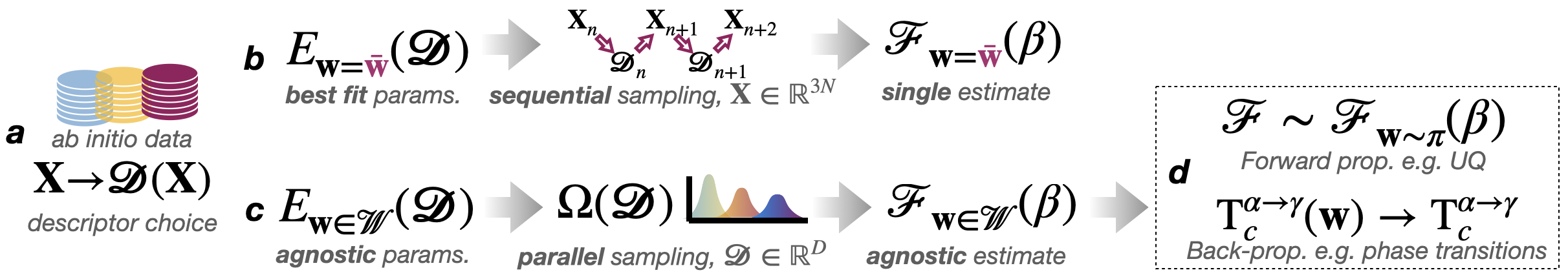}
    \caption{Model agnostic sampling with the descriptor density of states (D-DOS) for uncertainty quantification (UQ) and inverse design. 
    a) Modern atomic simulations choose some set of descriptor functions 
    to represent atomic structure (section \ref{sec:descriptors}) 
    and build interatomic potentials (MLIPs) that approximate some \textit{ab initio} dataset.
    b) Existing approaches determine a \textit{single} best fit of MLIP parameters 
    then execute sequential Monte Carlo sampling (section \ref{sec:FE}),
    potentially using descriptors to bias proposals\cite{tamagnone2024coarse,ahmad2022free,wirnsberger2022normalizing,noe2019boltzmann}. 
    The output is a free energy estimate for the \textit{specific} value of MLIP parameters employed.
    c) The D-DOS approach instead has \textit{agnostic} model parameters
    in some region $\w\in\mathcal{W}$, estimating the D-DOS
    in a parallelized score-matching scheme with around $10\times$ efficiency (section \ref{sec:numerics}). 
    The output is an \textit{agnostic} estimator which returns free energies and gradients
    for any parameter value $\w\in\mathcal{W}$.
    d) D-DOS enables rapid UQ in forward propagation and, uniquely,
    inverse design in back-propagation (section \ref{sec:experiments}) 
    to fine-tune e.g. phase transition temperatures. 
    In d), $\sim$ indicates `distributed as'.\label{fig:overview-0}}
\end{figure*}
However, all current approaches still require \textit{a priori} specification of MLIP parameters before any sampling is performed.
The restriction to \textit{specific} model parameters 
significantly complicates exploration of how MLIP parameters influence finite temperature properties, which has gained increasing interest for 
both uncertainty quantification\cite{swinburne2025,imbalzano2021uncertainty,musil2019fast,best_confirmal_2024,Zhu2023} 
in forward-propagation and an array of inverse design goals in back-propagation\cite{thaler2022deep,lopanitsyna2023modeling},
such as targeted fine-tuning\cite{grigorev2023calculation,maliyov2024exploringparameterdependenceatomic}.\\

In this paper, we propose a completely new approach 
for model-\textit{agnostic} sampling, introducing the 
{descriptor density of states} (D-DOS) $\Omega(\tD)$, a generalization of the energy DOS. 
Our main result is the D-DOS entropy (log density) can be efficiently estimated 
via score matching to yield an accurate free energy estimator
without \textit{a priori} specification of model parameters.
Numerical experiments demonstrate forward propagation for UQ and back propagation for
inverse fine-tuning, to our knowledge a first demonstration for free energies and phase boundaries.
An overview of the D-DOS approach in relation to existing schemes is illustrated in figure \ref{fig:overview-0}.

The paper is structured as follows. After reviewing existing free energy estimation 
schemes in section \ref{sec:FE}, we discuss descriptor-based MLIPs in section \ref{sec:descriptors}, 
where we show how a broad class of MLIPs\cite{Shapeev_MTP,Thompson_snap_2015,lysogorskiy2021performant,goryaeva2021,Nguyen2023,batatia2023foundation,perez2025uncertaintyquantificationmisspecifiedmachine} can be expressed as a linear 
sum of descriptor features. Section \ref{sec:general} then introduces the D-DOS, 
showing how the exponential divergence of any density of states 
expression\cite{wang2001efficient,partay2021nested} via a {conditional} scheme 
in close analogy to free energy perturbation. While general to any 
descriptor-based MLIP, in section \ref{sec:Z-laplace} we show that for MLIPs 
linear in descriptor features, the free energy can be exactly expressed as the Legendre-Fenchel conjugate of the descriptor entropy, avoiding high-dimensional integration through Laplace's method in the limit $N\to\infty$. 
We show the descriptor entropy has a maximum of zero, meaning we can 
integrate the entropy gradient estimated via score matching\cite{hyvarinen2005estimation},
detailed in section \ref{sec:SM}.

Our final result is a lightweight, differentiable, 
model-agnostic estimate that can be employed in forward or back propagation. 
In section \ref{sec:numerics}, we detail the numerical implementation of reference  
free energy calculations and our score matching scheme, for which an open source 
code is provided\cite{descriptordos}. Section \ref{sec:experiments} presents
extensive numerical experiments, demonstrating the ability of our approach to predict solid state free energies for a broad range of models, covering BCC and A15 phases of W and Mo, 
finding typical errors of 1-2 meV/atom against established free energy estimation schemes. 
We then provide a first exploration of inverse fine-tuning for free energy curves
focusing on the BCC and FCC phases of W and Fe, targeting the $\alpha\to\gamma$ transition temperature.

We discuss the many possible perspectives for our approach in the conclusions (section \ref{sec:conclusions}), including inverse design applications and extension to foundational MLIPs\cite{batzner2022,batatia2023foundation, Bochkarev2024}. As our approach allows practical implementation of density of states methods\cite{wang2001efficient,partay2021nested} to very high-dimensional systems, we anticipate application across the many areas of computational science that use linear feature models\cite{Marchand2023,follain2024nonparametric,Erhart2024}.
\section{Introductory example with the energy density of states}
As a guide to the reader, we first illustrate our approach by showing 
how access to the energy density of states $\Omega(U)$ allows for 
temperature-agnostic sampling, which this paper generalizes to 
model-agnostic sampling with $\Omega(\tD)$, the descriptor density of states (D-DOS). 
Leaving detailed definitions for later sections, with $\Omega(U)$ 
the classical partition function at ${\rm T}=1/({\rm k_B}\beta)$ writes
\begin{equation}
    Z(\beta)=\int_0^\infty \Omega(U)\exp(-\beta U){\rm d}U.
\end{equation}
For systems with local interactions $\Omega(U)$ can be cast 
in terms of an \textit{intensive} $\mathcal{S}(u)$ as $N\to\infty$, where $u=U/N$:
%We work with the  energy $u=U/N$ and will 
%express the energy DOS via an {intensive} entropy $\mathcal{S}(u)$:
\begin{equation}
    \Omega(U) \to \Omega_0 e^{N \mathcal{S}(u)},
    \quad
    Z(\beta) \to Z_0\int_0^\infty e^{N(\mathcal{S}(u)-\beta u)}du,
    \nonumber
\end{equation}
with $\Omega_0,Z_0$ constants. $Z(\beta)$ is dominated by the maximum of the integrand as $N\to\infty$,
allowing exact free energy evaluation via Laplace's method (appendix \ref{app:laplace}):
\begin{equation}
    \mathcal{F}(\beta)\equiv\lim_{N\to\infty}\frac{-\ln|Z(\beta)|}{\beta N}
    =
    \min_{u}u-\mathcal{S}(u)/\beta.
\end{equation}
We thus see that access to $\mathcal{S}(u)$ allows temperature agnostic estimation 
through minimization with  $-\beta\mathcal{F}(\beta)$ defined as 
the Legendre-Fenchel conjugate of $\mathcal{S}(u)$\cite{fenchel_conjugate_1949}.
This paper generalizes the above from $\Omega(U)$ to $\Omega(\tD)$, enabling model-agnostic MLIP sampling.\section{Thermodynamic sampling of atomic crystal models}
\label{sec:FE}
This section reviews standard results from classical statistical mechanics for a system
of $N$ atoms with specie ${\bf s}=[{\rm s}_1,\dots,{\rm s}_N]\in\mathbb{Z}^N$,
positions $\X=[{\bf x}_1,\dots,{\bf x}_N]\in\mathbb{R}^{N\times3}$ 
and momenta $\mom=[{\bf p}_1,\dots,{\bf p}_N]\in\mathbb{R}^{N\times3}$.
Atoms are confined to a periodic supercell ${\bf C}\in\mathbb{R}^{3\times3}$ 
with volume $V=|{\bf C}|$ (the determinant), such that scaled positions lie on the unit torus, 
i.e. $\X{\bf C}^{-1}\in\mathbb{T}^{N\times3}$. In anticipation of later 
results where we take the limit $N\to\infty$, we write the total energy 
$U(\X,\mom)$ as the sum of an \textit{intensive} potential and kinetic energy, 
i.e. 
\begin{equation}
    U({\bf X},\mom) \equiv N[{E}_{\w}(\X) + {K}(\mom)],
    \label{per-atom-energy}
\end{equation}
where ${K}(\mom)=(1/N)\sum_{i=1}^N{\rm p}^2_i/(2m_i)$ 
and dependence on ${\bf s}$ is contained in the potential energy function ${E}_\w(\X)$
by model parameters $\w$, the focus of this paper, treated in section \ref{sec:descriptors}.
To express the supercell in an intensive form we define the supercell per atom $\C$
through ${\bf C}={\bf N}\C$, where ${\bf N}={\rm Diag}(N_x,N_y,N_z)$, 
such that $|{\bf N}|=N$ and the volume per atom is given by $|\C|$.
The canonical (NVT) partition function at ${\rm T}=1/({\rm k_B}\beta)$ then writes
\begin{align}
Z^N_\w(\beta,\C)
%&\equiv 
%\int_{\mathbb{R}^{6N}}\exp(-N\beta[{K}(\mom)+ E_\w(\X)])
%{\rm d}\mom{\rm d}\X,\\
&\equiv 
    %\frac{1}{h^{3N}}
    \lambda_0(\beta)^{-3N}
    %\frac{\sqrt{{\beta}/{\tilde{m}}}^{3N}}{h^{3N}}
    %\sqrt{\frac{\beta}{\tilde{m}}}^{3N}
    \int_{\mathbb{R}^{3N}}\exp[-N\beta E_\w(\X)]
    {\rm d}\X,
    \label{nvt_Z}
\end{align}
where
$\lambda_0(\beta) = \hbar\sqrt{2\pi\beta/m}$ is the thermal De Broglie wavelength\cite{Book_Frenkel} and 
${m}^N\equiv\prod_{i=1}^Nm_i$.
%and the integral over momenta is performed analytically.
The NVT free energy per atom is defined in the thermodynamic limit $N\to\infty$:
\begin{equation}
    \mathcal{F}_\w(\beta,\C)
    \equiv 
    \lim_{N\to\infty}\frac{-1}{\beta N}\ln
    |Z^N_\w(\beta,\C)|.
    \label{limit_free_energy}
\end{equation}
In practice, the integral over atomic configuration space 
in (\ref{nvt_Z}) is dominated by contributions from 
some set of \textit{phases} $\mathcal{P}=\{$bcc, fcc, hcp, liquid, $\dots\}$,
such that
\begin{equation}
    Z^N_\w(\beta,\C)=\sum_{p\in\mathcal{P}} Z^N_\w(\beta,\C,p),
\end{equation}
where each term $Z^N_\w(\beta,\C,p)$,
is an integral over (disjoint) partitions of configuration space, with 
corresponding phase free energy $\mathcal{F}_\w(\beta,\C,p)$, 
defined as in (\ref{limit_free_energy}). It is simple to show that as
$N\to\infty$ the NVT free energy is dominated by a single phase
\begin{equation}
    p_\w^*(\beta,\C) = \arg \min_{p\in\mathcal{P}}\mathcal{F}_\w(\beta,\C,p),
\end{equation}
as $\mathcal{F}_\w(\beta,\C) = \min_{p\in\mathcal{P}}\mathcal{F}_\w(\beta,\C,p)$.
Similarly, the NPT free energy of a phase $p$ is obtained by minimizing 
$\mathcal{F}_\w(\beta,\C,p)$ at under some constant external 
stress ${\bm\sigma}$ (i.e. isotropic pressure ${\bm\sigma}=(P/3)\mathbb{I}_3$), giving
\begin{equation}
    \mathcal{G}_\w(\beta,{\bm\sigma},p) \equiv \min_\C
    \mathcal{F}_\w(\beta,\C,p) - {\rm Tr}({\bm\sigma}^\top\C),
    \label{gibbsF}
\end{equation}
%such that $\partial_\C\mathcal{F}_\w(\beta,\C^*,p) \equiv {\bm\sigma}$.
It is clear that estimation of $\mathcal{F}_\w(\beta,\C,p)$ for general $\beta,\C$
is sufficient to estimate $\mathcal{G}_\w(\beta,{\bm\sigma},p)$, 
giving the stable phase at some temperature and pressure as
\begin{equation}
    p^*_\w(\beta,{\bm\sigma}) = \arg\min_{p\in\mathcal{P}}\mathcal{G}_\w(\beta,{\bm\sigma},p),
    \label{phase_stability}
\end{equation}
where the $\w$ subscript emphasizes the dependence of $p^*$ on the parameters of the interatomic potential.\\

In this paper, we will focus on the set of crystalline phases $\mathcal{P}_s\subset\mathcal{P}$, 
whose configuration space is defined as the set of (potentially large) vibrations around some
lattice structure $\X_p^0$, $p\in\mathcal{P}_s$. 
Extension of the present approach
to liquid phases is discussed in \ref{sec:conclusions}.
\subsection{Thermodynamic integration}\label{sec:TI}
Accurate calculation of phase stability requires converging per-atom 
free energy differences between phases to within a few meV/atom at any given temperature 
and pressure to allow determination of (\ref{phase_stability}). 
Accurate determination of phase transitions, where free energy differences are formally zero\cite{zhu2017efficient,grabowski2019ab,Zhong2023,zhu2024accelerating,menon2024electrons},
thus requires tight convergence of any estimator. The stringent accuracy requirement 
has led to the development of sampling techniques to reduce the number of samples required for convergence~\cite{kirkwood1935statistical,Book_Frenkel,rickman2002free}. 
In all cases, the starting point is some atomic energy 
function $E_0({\bf X})$, whose corresponding phase free energy 
$\mathcal{F}_0(\beta,\C,p)$ is known either through 
tabulation, or analytically if $E_0({\bf X})$ is harmonic\cite{kirkwood1935statistical}. We can thus define $\Delta E_\w(\X)=E_\w(\X)-E_0(\X)$ as the energy difference (per-atom)
between the target and reference systems, with a free energy difference $\Delta\mathcal{F}_\w(\beta,\C,p)$. 
Thermodynamic integration (TI) is a stratified sampling scheme over 
$E_\eta({\bf X})=E_0({\bf X}) + \eta \Delta E_{\bf w}({\bf X})$
for $\eta\in[0,1]$. 
%\mcm{There are many schemes beyond linear in \( \eta \) that are mostly used when non-equilibrium sampling is performed in a Jarzynski-like manner, e.g., the approach adopted in the paper we recently discussed by Sarath Menon, Jutta Rogal, Drautz, etc.} 
Denoting equilibrium averages by
$\langle\dots\rangle_{\eta}$, we obtain 
\begin{align}
    \Delta\mathcal{F}_{\bf w}(\beta,\C,p)
    &=
    \int_0^1\langle\Delta E_{\bf w}({\bf X})\rangle_{\eta}\,{\rm d}\eta.
    \label{TI}
\end{align}
Sampling efficiency often requires constraint functions 
or resetting to prevent trajectories escaping the metastable basin
of a given phase, as discussed in section \ref{sec:TI-sampling}.
In general, the larger the value of $\Delta\mathcal{F}_{\bf w}$, 
the finer the integration scheme over $\eta$ and the more samples 
will be required for convergence \cite{rousset2010free, Comer2014}.
\subsection{Free energy perturbation}\label{sec:FEP}
Typically used as a complement to thermodynamic integration, if the difference 
$N\Delta E_\w(\X)$ is as small as $10/\beta$, corresponding to 
at most 10 meV/atom at 1000 K for solid state systems ($N\simeq100$), 
we can also use free energy perturbation (FEP) to estimate the free energy difference\cite{rousset2010free,grabowski2019ab,castellano2022b}.
Using the definition of the free energy $\mathcal{F}_\w(\beta,\C)$ and $\langle\dots\rangle_{\eta}$ at $\eta=0$, it is simple to show that
\begin{align}
    \Delta\mathcal{F}_{\bf w}(\beta,\C,p)
    &=
    -(1/N\beta)
    \ln\langle \exp[-N\beta\Delta E_{\bf w}(\X)]\rangle_{0}.
    \nonumber
    %label{FEP}
\end{align}
In practice, the logarithmic expectation is expressed as a cumulant expansion\cite{zwanzig_fep_1954, rousset2010free, castellano2024machinelearningassistedcanonical} 
for increased numerical stability,
writing 
\begin{equation}
    \Delta\mathcal{F}_{\bf w}(\beta,\C,p)
    =
    \langle{\Delta E_{\bf w}}\rangle_{0}
    -\frac{N\beta}{2}
    \langle(\delta\Delta E_{\bf w})^2\rangle_{0}
    +
    \dots,
    \label{cumulant_expansion_FEP}
\end{equation}
where $\delta\Delta E_{\bf w}=\Delta E_{\bf w}-\langle{\Delta E_{\bf w}}\rangle_{0}$,
and the expansion continues, in principle, to all orders. 
While (\ref{cumulant_expansion_FEP})
gives an expression for the free energy in terms of samples generated
solely with a reference potential, in a practical setting we require 
free energy differences to be very small to allow for convergence. 
Equation (\ref{cumulant_expansion_FEP}) can be shown to be an upper bound to the estimated free energy difference\cite{castellano2022b} and as such can be used as a 
convergence measure for a well-chosen reference potential. 
In this setting, we typically have $\Delta\mathcal{F}_{\bf w}<1$ meV/atom (table \ref{tab:cost_comp}).
\subsection{Adiabatic Switching}\label{sec:AS}
In addition to the above methods which employ equilibrium averages,
the adiabatic switching\cite{ adjanor_free_2006, de1999optimized,freitas2016nonequilibrium,menon2024electrons} method estimates free energy differences using the well-known Jarzynski equality \cite{jarzynski1997nonequilibrium}. The 
adiabatic switching equality can be written\cite{freitas2016nonequilibrium}
\begin{align}
    \Delta\mathcal{F}_{\bf w}(\beta,\C,p)
    &=
    \frac{1}{2}[
    \langle W^{\textrm{irr}}\rangle_{0\to1}-
    \langle W^{\textrm{irr}}\rangle_{1\to0}
    ],
\end{align}
where $W^{\textrm{irr}}$ is the irreversible work along a thermodynamic path (in the above $\eta$ is implied, though it is also possible to use the temperature) and $\langle\dots\rangle_{0\to1}$ indicates an ensemble average of around $10-30$ simulations. The key quantity is 
the so-called `switching time', i.e. the rate at which the thermodynamic path is 
traversed. For solid-state free energies one typically progresses 
along the path in $\mathcal{O}(10)$ increments of $\mathcal{O}(10-100)$ ps\cite{menon2024electrons}, thus requiring around 
$10^{7-8}$ force calls per temperature. In this setting, we can 
target similar free energy differences to thermodynamic integration, 
i.e. $\mathcal{O}(100)$ meV/atom at 1000 K. The computational 
costs of the above methods and the present D-DOS approach 
is discussed in section \ref{sec:numerics}, and summarized in
table \ref{tab:cost_comp}.

\section{Descriptor-based interatomic potentials}
\label{sec:descriptors}
The search for efficient and accurate interatomic potentials
is a central goal of computational materials science\cite{deringer2019machine}.
For systems without long-range interactions, 
the intensive energy in (\ref{per-atom-energy}) is modeled as
a sum of functions over local atomic environments
\begin{equation}
    E_\w(\X)=\frac{1}{N}\sum_{i=1}^N E^1_\w({\bf R}_i,{\bf s}_i),
    \label{per-atom-E}
\end{equation}
where ${\bf R}_i=[\dots,{\bf r}_{ij},\dots]$ is a set of vectors from $i$ to 
all neighbors $j$ within some cutoff $r_c$, ${\bf s}_i=[\dots,{\rm s}_{j},\dots]$ are the 
corresponding chemical species (along with ${\rm s}_i$ of $i$) 
and $\w$ are learnable model parameters. Modern interatomic potentials typically represent
local atomic environments through scalar-valued and dimensionless (i.e. unit-free) descriptor vectors
\begin{equation}
    \D_i \equiv \hat{\D}({\bf R}_i,{\bf s}_i,{\bf h})\in\mathbb{R}^d
\end{equation}
where ${\bf h}\in\mathbb{R}^H$ are a set of hyperparameters for encoding atomic species and all adjustable parameters in $\hat{\D}$, and $d$ is the descriptor dimension. A general descriptor-based 
interatomic potential then has total energy 
\begin{equation}
    E_\w(\D_1,\dots,\D_N)=\frac{1}{N}\sum_{i=1}^N E^1_\w(\D_i),\label{descriptorE}
\end{equation}
where we suppress the chemical vectors ${\bf s}_i$ as they are considered encoded in the $\D_i$. For systems with $M$ atomic species $d$ formally increases combinatorially with $M$, but in practice sparsification schemes give 
$d=\mathcal{O}(M)$\cite{darby2023tensor,Thompson_snap_2015}. 
%Hyperparameters ${\bf h}$ are typically low rank ($H=\mathcal{O}(10)$) for descriptor-based models and are considered fixed in the following. 
In this work, we consider potentials of the general linear form, with parameters $\w\in\mathbb{R}^D$,
\begin{equation}
    E^L_\w(\tD) \equiv \w^\top\tD,
    \quad 
    \tD \equiv \frac{1}{N}\sum_{i=1}^N
    \hat{\bphi}(\D_i)\in\mathbb{R}^D,
    \label{linearE}
\end{equation}
where $\hat{\bphi}(\D_i)=[\hat{\phi}_1(\D_i),\dots,\hat{\phi}_D(\D_i)]$
is a $D$-dimensional featurization of $\D_i\in\mathbb{R}^d$ and 
the $L$ superscript denotes the linear form. Importantly, $\tD$ is independent of parameters $\w$, and the dimension $D$ is typically larger than $d$ but \textit{intensive}, i.e. independent of $N$, which is required to apply Laplace's method in section \ref{sec:Z-laplace}. 
\subsection{Admissible interatomic potentials}\label{sec:mlip-types}
A wide variety of interatomic potentials can be cast into the general 
linear form (\ref{linearE}). Clearly, these include the broad class of 
linear-in-descriptor (LML) models, including \texttt{MTP}\cite{Shapeev_MTP}, \texttt{ACFS} \cite{behler2011, behler2016},
\texttt{SNAP}\cite{Thompson_snap_2015}, \texttt{SOAP} \cite{bartok_2013, darby_compressing_soap_2022, darby2023tensor}, \texttt{ACE}\cite{drautz_atomic_2019, drautz_atomic_2020, lysogorskiy2021performant},
\texttt{MILADY}\cite{goryaeva2021, dezaphie_designing_2025} and \texttt{POD}\cite{Nguyen2023} descriptors, where
\begin{equation}
    \hat{\bphi}_{\tt{LML}}(\D_i) = \D_i\in\mathbb{R}^d.\label{LML-DESC}
\end{equation}
LML models can reach extremely high ($<2$ meV/atom) accuracy to \textit{ab initio} data\cite{castellano2024machinelearningassistedcanonical}, with robust UQ\cite{swinburne2025} and often excellent dynamical stability, essential for thermodynamic sampling\cite{zhu2017efficient,grabowski2019ab,Zhong2023,zhu2024accelerating,menon2024electrons}. Polynomial or kernel featurizations are regularly used to increase flexibility, e.g. \texttt{qSNAP}\cite{Rohskopf2023}, \texttt{PiP}\cite{allen_atomic_2021} 
\texttt{GAP}\cite{bartok_2015, deringer2019machine}, n-body kernels \cite{glielmo_accurate_2017, glielmo_efficient_k2b_2018, vandermause_flare_2020, xie_flare_2021}, kernel\cite{Zhong2023} etc.
For example, with \texttt{qSNAP} we have the quadratic featurization
\begin{equation}
    \hat{\bphi}_{\tt{qSNAP}}(\D_i)=
    \D_i\oplus{\tt{vech}}(\D_i\otimes\D_i)
    \in\mathbb{R}^{d(d+3)/2},
\end{equation}
i.e. $D=d(d+3)/2$ as ${\tt{vech}}(\D_i\otimes\D_i)\in\mathbb{R}^{d(d+1)/2}$ returns a vector from the quadratic product $\D_i\otimes\D_i\in\mathbb{R}^{d\times d}$.\\

The generalized linear form can encompass more complex models
if we only consider a subset of parameters adjustable, an approach adopted 
when fine-tuning recent message-passing neural network (MPNN) models\cite{batzner2022, 
musaelian_allegro_2023, batatia2022mace,batatia2023foundation,Bochkarev2024,cheng2024cartesian,batatia2025design}.
In this case, we consider inputs of the MPNN readout layer $h(\D_i)$ 
as descriptors $\D_i$, then express the readout layer as a general linear model 
of the desired form (\ref{linearE}). For example, in the 
\texttt{MACE} MPNN architecture\cite{batatia2022mace}, the readout layer is a 
sum of a linear function and a one layer neural network $f(\D_i)$, giving
the featurization
\begin{equation}
    \hat{\bphi}_{\tt{MACE}}(\D_i)=\D_i\oplus f(\D_i)\in\mathbb{R}^{d+1},
\end{equation}
i.e. $D=d+1$ as we treat the neural network as a fixed feature function, such that 
adjustable parameters $\w$
cover all linear-in-descriptor terms and a scalar prefactor on the neural network. 
Recent work has shown this allows uncertainty quantification schemes for linear models\cite{swinburne2025} to be applied to the \texttt{MACE-MPA-0} foundation model\cite{batatia2023foundation}, successfully bounding prediction errors across the materials project database\cite{perez2025uncertaintyquantificationmisspecifiedmachine}.
More generally, one could also aim to learn a set of descriptor feature functions 
$\hat{\bphi}(\D_i)$ \textit{and} a set of parameter feature functions $\hat{\bpsi}(\w)$
to minimize the loss $L(\T)=\sum_i\|\hat{\bpsi}(\w)^\top\hat{\bphi}(\D_i)-E_\w(\D_i)\|^2$
to some $E_\w(\D_i)$, giving a generalized linear feature model under $\w\to\hat{\bpsi}(\w)$,
a direction we leave for future research.
As a result, while the numerical applications of this paper (section \ref{sec:numerics}) focus on linear MLIPs with \texttt{SNAP} descriptors, i.e. equation (\ref{LML-DESC}), 
the D-DOS approach opens many broader perspectives which are discussed further in section \ref{sec:conclusions}.

\section{The descriptor density of states}
\label{sec:general}
%
%
%\subsection{Descriptor density of states}\label{sec:gen_dos}
The NVT free energy is defined in equation (\ref{limit_free_energy}). Suppressing dependence on $p$ and $\C$ for clarity, the free energy 
for the generalized linear MLIPs (\ref{linearE}) can be written
\begin{equation}
    \mathcal{F}^L_\w(\beta)
    =
    \lim_{N\to\infty}
    \frac{-1}{N\beta}\ln
    \left|
    \lambda_0^{-3N}(\beta)
    \int_{\mathbb{R}^D}
    e^{-N\beta \w\cdot\tD}
    \Omega(\tD)
    {\rm d}\tD
    \right|,\nonumber
\end{equation}
where $\Omega(\tD)$ is the \textit{descriptor density of states} (D-DOS)
\begin{equation}
    \Omega(\tD)
    \equiv 
    \int_{\mathbb{R}^{3N}}
    \delta\left(
    \tD-\sum_{i=1}^N\hat{\bphi}(\D_i({\bf X}))/N
    \right) 
    {\rm d}\X.
    \label{DDOS}
\end{equation}
Estimation of $\Omega(\tD)$ would allow prediction of $\mathcal{F}_\w(\beta)$ for \textit{any value} of the potential parameters $\w$, our central goal. 
However, this requires overcoming two significant numerical issues. 
First, $\Omega(\tD)$ is ill-conditioned, with an integral that grows exponentially with $N$:
\begin{equation}
    \int_{\mathbb{R}^D}\Omega(\tD)
    {\rm d}\tD=
    V^N.\label{WL}
\end{equation}
This divergence is common to all density of states, e.g. 
$\int_{\mathbb{R}}\Omega(E){\rm d}E=V^N$, and severely limits the application of Wang-Landau\cite{wang2001efficient} or nested sampling\cite{partay2021nested} at large $N$. 
Secondly, the global feature vector $\tD\in\mathbb{R}^{D}$ has a very high dimension, typically $\mathcal{O}(100-1000)$ and thus (\ref{WL}) cannot be integrated through direct quadrature, while Monte Carlo estimation is generally slow to converge and cannot give reliable gradient information. 
The central contributions of this paper are strategies to overcome these two issues. 
Section \ref{sec:CDDOS} contains the $V^N$ divergence via a conditional scheme,
and section \ref{sec:Z-laplace} shows how Laplace's method (appendix \ref{app:laplace})
allows us to avoid high-dimensional integration.
%, in close connection to free energy perturbation (\ref{sec:FEP}).
%while  allows us to avoid high-dimensional integration, casting
%free energy evaluation as a convex minimization in the limit $N\to\infty$.
%
%
%
\subsection{The conditional descriptor density of states}\label{sec:CDDOS}
To control the $V^N$ divergence of $\Omega(\tD)$, we introduce the 
\textit{conditional} descriptor density of states (CD-DOS)
\begin{align}
    \Omega(\tD|\alpha)
    \equiv 
    &\int_{\mathbb{R}^{3N}}
    \frac{\delta\left(\hat{\alpha}(\X)-\alpha\right)}{\Omega(\alpha)}
    \nonumber
    \\
    &
    \times
    \delta\left(
    \tD-\sum_{i=1}^N\hat{\bphi}(\D_i({\bf X}))/N
    \right)
    %\prod_{j}\delta(\hat{\D}_j(\X)-\D_j)
    {\rm d}\X,
    \label{cdos}
\end{align}
where $\hat{\alpha}(\X)$ is the dimensionless function 
\begin{equation}
    \hat{\alpha}(\X)
    \equiv 
    \ln\left|{{E}_0({\bf X})}/{\rm U}_0\right|,
    \label{isosurface-X}
\end{equation}
${\rm U}_0$ is a user-defined energy scale and ${E}_0({\bf X})\geq0$ 
is some intensive reference potential energy, as in equation (\ref{per-atom-energy}).
In section \ref{sec:gen_alpha} we 
detail how equation (\ref{isosurface-X}) can be generalized to a 
momentum-dependent $\hat{\alpha}(\X,\mom)$.
In either case, ${E}_0({\bf X})$ is chosen such that we can 
calculate, numerically or analytically, the isosurface volume
\begin{equation}
    \Omega(\alpha)\equiv\int_{\mathbb{R}^{3N}}
    \delta\left(
        \hat{\alpha}(\X)-\alpha
    \right)
    {\rm d}{\bf X},
    \label{volume}
\end{equation}
which contains the exponential divergence as 
$\int_{\mathbb{R}}\Omega(\alpha){\rm d}\alpha=V^N$.
% \Omega_0 shouldn't be there !
%The isosurface volume $\Omega(\alpha)$
%
%\mcm{Where, $\Omega_0$
%encapsulates the contribution from the Jacobian and the measure change induced by the %transformation from the full coordinates 
%$\X$  to the new variables (one of which is $\alpha$)}
%
The crucial advantage of this conditional form is that 
estimation of $\Omega(\tD|\alpha)$ is then
much simpler, as it is normalized by construction:
\begin{equation}
    \int_{\mathbb{R}^D}\Omega(\tD|\alpha){\rm d}\tD
    =\frac{\Omega(\alpha)}{\Omega(\alpha)}
    = 1.
    \label{condOm_norm}
\end{equation}
This normalization allows us to employ density estimation techniques 
such as score matching\cite{hyvarinen2005estimation}.
Equation (\ref{condOm_norm}) shows $\Omega(\tD|\alpha)$ is the probability density function 
of $\tD$ on the isosurface $\hat{\alpha}(\X)=\alpha$. The full D-DOS $\Omega(\tD)$
is then formally recovered through integration against $\alpha$:
\begin{equation}
    \Omega(\tD)
    =
    \int_{\mathbb{R}}
    \Omega(\tD|\alpha)
    \Omega(\alpha){\rm d}\alpha.
    \label{ddos-recon}
\end{equation}
Equation (\ref{ddos-recon}) emphasizes that our goal is to 
decompose the high-dimensional configuration space into 
a foliation of isosurfaces $\hat{\alpha}(\X)=\alpha$ 
where we expect $\Omega(\tD|\alpha)$ to be
tractable for density estimation. The generalization of
$\hat{\alpha}(\X)$ to include momentum dependence is
discussed in section \ref{sec:gen_alpha} and general 
considerations for designing optimal $\hat{\alpha}(\X)$ 
are discussed in section \ref{sec:laplace-error} 

\subsection{The isosurface and descriptor entropies}

The free energy $\mathcal{F}_\w(\beta)$, equation (\ref{limit_free_energy}), 
is proportional to the intensive logarithm of the 
partition function $Z_\w(\beta)$, i.e. $\mathcal{F}_\w(\beta)=(-1/N\beta)\ln|Z_\w(\beta)|$.
To estimate free energies, it is thus natural to define intensive \textit{entropies} 
of the isosurface volume $\Omega(\alpha)$ and CD-DOS $\Omega(\tD|\alpha)$. 
As we use $\beta=1/({\rm k_BT})$ rather than $\rm T$, 
we omit factors of ${\rm k_B}$ such that entropies are dimensionless.

We first define the intensive isosurface entropy
\begin{equation}
    \mathcal{S}_0(\alpha)\equiv \lim_{N\to\infty}
    \frac{1}{N}\ln|\Omega(\alpha)/V^N_0|.
    \label{entropy}
\end{equation}
The term $V_0$ ensures $\mathcal{S}_0(\alpha)$ is dimensionless;
with $\hat{\alpha}(\X)$ we have $V_0=\lambda^3_0(\beta)$, while 
with a momentum-dependent $\hat{\alpha}(\X,\mom)$, discussed in \ref{sec:gen_alpha},
we have $V_0=h^3$. It is clear that $\mathcal{S}_0(\alpha)$ 
is a measure of the configurational entropy per atom of $N$ 
independent atoms confined to the isosurface. The CD-DOS
$\Omega(\tD|\alpha)$, equation (\ref{cdos}), has a natural entropy definition, 
the intensive log density
\begin{equation}
    \mathcal{S}({\tD}|\alpha)\equiv
    \lim_{N\to\infty}
    \frac{1}{N}\ln\Omega({\tD}|\alpha).
    \label{descriptor-entropy}
\end{equation}
The CD-DOS entropy $\mathcal{S}(\tD | \alpha)$ measures the proportion of the 
isosurface phase space volume that has a global descriptor vector $\tD$, meaning 
descriptor values with larger $\mathcal{S}({\tD}|\alpha)$ 
are more likely to be observed under unbiased isosurface sampling.
Furthermore, $\mathcal{S}({\tD}|\alpha)$ has two properties which greatly 
facilitate free energy estimation. 
In appendix \ref{app:intensive}, we show that as the per-atom descriptor features 
$\bphi(\D_i)\in\mathbb{R}^D$ depend only on the local environment of atom $i$,
the CD-DOS entropy $\mathcal{S}({\tD}|\alpha)$ is intensive ($N$-independent).
Secondly, as $\Omega(\tD|\alpha)$ is normalized,  
application of Laplace's method (appendix \ref{app:laplace}) 
gives a very useful result that fixes the maximum of $\mathcal{S}({\tD}|\alpha)$:
\begin{equation}
    \lim_{N\to\infty}\frac{1}{N}\ln\left|\int\Omega({\tD}|\alpha){\rm d}\tD\right|
    =
    \max_{\tD\in\mathbb{R}^D} \mathcal{S}({\tD}|\alpha)
    =0.\label{szero}
\end{equation}
In section \ref{sec:Z-laplace} we show that free energy estimation for
linear‐in‐descriptor MLIPs reduces to a minimization 
over $\mathbf{w}^\top \tD - \beta^{-1}[\mathcal{S}(\tD|\alpha)+\mathcal{S}_0(\alpha)]$,
avoiding high-dimensional integration. Moreover, section \ref{sec:SM} shows 
$\mathcal{S}(\tD|\alpha)$ can be approximated through score matching of $\nabla_{\tD}\mathcal{S}(\tD | \alpha)$, where (\ref{szero}) fixes the subsequent constant of integration to give an absolute
estimate of $\mathcal{S}({\tD}|\alpha)$. 
\subsection{Forms of the isosurface function}\label{sec:gen_alpha}
As discussed above, free energy estimation will require access to 
$\mathcal{S}_0(\alpha)$ and a means to generate samples on the isosurface 
$\hat{\alpha}(\X)=\alpha$. For harmonic reference potentials 
$S_0(\alpha)$ is given analytically; the isosurface function writes
\begin{equation}
\hat{\alpha}(\X)
\equiv
\ln\left|{[\X-\X_0]^\top{\bf H}[\X-\X_0]}/({2N\rm U_0})\right|,\label{isosurface-H}
\end{equation}
where the Hessian ${\bf H}$ has 3$N$-3 positive eigenmodes 
and $\X_0$ is the lattice structure.
As detailed in appendix \ref{app:isosurface_harmonic}, 
sampling $\hat{\alpha}({\bf X})=\alpha$ reduces to generating 
random unit vectors in $\mathbb{R}^{3N-3}$, 
while the isosurface entropy (\ref{entropy}) reads
\begin{align}
   \mathcal{S}_0(\alpha)\equiv
   S_0 + 3\alpha/2,
   \quad
   V_0=\lambda^3_0(\beta).
   \label{lim_dos}
\end{align}
In appendix \ref{app:isosurface_harmonic}, we show 
the constant $S_0$ is given by $S_0=3/2 + 3/2 \ln|2\beta{\rm U}_0/3|-\beta\mathcal{F}_0(\beta)$, 
where $\mathcal{F}_0(\beta)$ is the familiar free energy per-atom of an atomic 
system governed by the harmonic potential $E_0(\X)$.

To go beyond harmonic reference models to arbitrary $E_0(\X)$,
we generalize (\ref{isosurface-X}) to accommodate the intensive 
kinetic energy ${K}(\mom)=(1/N)\sum_{i=1}^N{\rm p}^2_i/(2m_i)$
from (\ref{per-atom-energy}), such that the isosurface value is the log total energy
\begin{equation}
    \hat{\alpha}(\X,\mom)
    \equiv 
    \ln\left|[{K}({\bf P}) + E_0({\bf X})]/{\rm U}_0\right|.
    \label{isosurface-P}
\end{equation}
Isosurface sampling then reduces to running microcanonical (NVE) dynamics,
in close connection with Hamiltonian Monte Carlo\cite{betancourt2017conceptual,livingstone2019kinetic}.
In appendix \ref{app:isosurface_momentum},
we show the NVT free energy $\mathcal{F}_0(\beta)$ of the 
reference system can be expressed as $\mathcal{F}_0(\beta)=\mathcal{U}_0(\beta)-\mathcal{S}_0(\alpha_\beta)$, 
where $\mathcal{U}_0(\beta)$ is the 
internal energy per atom and $\alpha_\beta\equiv\ln|\mathcal{U}_0(\beta)/{\rm U}_0|$.\\
As a result, with a momentum-dependent isosurface (\ref{isosurface-P}) 
the isosurface entropy (\ref{entropy}) is simply the difference between 
the reference system's free and internal energies:
\begin{align}
    \mathcal{S}_0(\alpha)=
    \beta_\alpha\left[\mathcal{U}_0(\beta_\alpha)-\mathcal{F}_0(\beta_\alpha)\right],
    \quad
    V_0(\alpha)=h^3,
    \label{lim_dos-F}
\end{align}
where $\beta_\alpha$ is defined through
$\mathcal{U}_0(\beta_\alpha)\equiv{\rm U}_0\exp(\alpha)$,
which will have a unique solution when $\mathcal{U}_0(\beta_\alpha)$ 
is monotonic. In practice, the
internal and free energies $\mathcal{U}_0(\beta)$ and 
$\mathcal{F}_0(\beta)$ are estimated via thermodynamic 
sampling (\ref{sec:FE}) over a range of $\beta$, interpolating 
with $\alpha\equiv\ln|\mathcal{U}_0(\beta)/{\rm U}_0|$ to 
estimate $\mathcal{S}_0(\alpha)$.\\
The final modification, 
detailed in appendix \ref{app:isosurface_momentum}, is to augment the descriptor 
vector $\tD$, concatenating the kinetic energy 
as a scalar momentum descriptor $K={K}(\mom)$
\begin{equation}
    \tD\to\tD\oplus K,\quad \w\to\w\oplus1,
    \label{momentum_descriptor}
\end{equation}
meaning $\w^\top\tD$ now returns the total energy rather than the potential energy.\\

In conclusion, sampling schemes can thus use $\hat{\alpha}(\X)$ 
with a harmonic reference potential, where $\mathcal{S}_0(\alpha)$ is given analytically, 
or $\hat{\alpha}(\X,\mom)$ with any reference potential, where $\mathcal{S}_0(\alpha)$ 
determined via thermodynamic sampling. All theoretical results below can use 
either $\mathcal{S}_0(\alpha)$; use of both are demonstrated for solid phases in 
section \ref{sec:influence-score}. A forthcoming study will apply the momentum-dependent 
formalism to liquid phases and melting transitions. temperature prediction.

\section{Free energy evaluation with Laplace's method}\label{sec:Z-laplace}
Laplace's method, or steepest descents\cite{wong2001asymptotic}, is a common 
technique for evaluating the limiting form of integrals of exponential 
functions. Consider a twice differentiable function $f({\bf x}) : \mathbb{R}^p \to \mathbb{R} $ 
that has a unique minimum in $\mathbb{R}^p$ and is \textit{intensive}, i.e. independent of $N$.
Under some mild technical conditions, discussed in appendix \ref{app:laplace}, 
Laplace's method implies the limit
\begin{equation}
    \lim_{N\to\infty}
    \frac{-1}{N}\ln
    \int_{\mathbb{R}^p}
    \exp[-Nf({\bf x})]{\rm d}{\bf x}
    =
    \min_{{\bf x}\in\mathbb{R}^p}
    f({\bf x})
    .
    \label{laplace}
\end{equation}
Application of (\ref{laplace}) with $p=D$ and $p=1$ for integrals over $\tD$ and $\alpha$
is our primary device to avoid integration in free energy estimation. 
Using the definition of the CD-DOS entropy $\mathcal{S}({\tD}|\alpha)$, 
equation (\ref{descriptor-entropy}), shown to be intensive in appendix \ref{app:intensive},
application of (\ref{laplace}) implies that 
\begin{align}
    \mathcal{F}^L_{\w}(\beta|\alpha)
    &\equiv
    \lim_{N\to\infty}
    \frac{-1}{N\beta}
    \ln
    \left|
    \int_{\mathbb{R}^D}
    e^{-N\beta \w^\top\tD}
    \Omega(\tD|\alpha)
    {\rm d}\tD
    \right|,
    \nonumber\\
    &=
    \min_{\tD\in\mathbb{R}^D}
    \left(
    \w^\top\tD-\mathcal{S}({\tD}|\alpha)/\beta
    \right),
    \label{condF_final}
\end{align}
%\mcmc{It is not better to use parenthesis after $\min \left( \right)$ ? Not only here ...  } 
meaning $-\beta\mathcal{F}^L_{\w}(\beta|\alpha)$ is the Legendre–Fenchel \cite{fenchel_conjugate_1949}
conjugate of the entropy $\mathcal{S}({\tD}|\alpha)$ for linear MLIPs. 
It is clear that the conditional free energy (\ref{condF_final}) has a close connection to 
the cumulant expansion in free energy perturbation 
\cite{castellano2024machinelearningassistedcanonical}, equation (\ref{cumulant_expansion_FEP}), 
a point we discuss further in section \ref{sec:SM-FEP}. 
We thus obtain a final free energy expression
\begin{align}
    \mathcal{F}^L_{\w}(\beta)
    &\equiv 
    \lim_{N\to\infty}
    \frac{-1}{N\beta}
    \ln
    \int_{\mathbb{R}}
    e^{N[\mathcal{S}_0(\alpha)-
    %\beta\mathcal{K}_0(\beta)-
    \beta\mathcal{F}_{\w}(\beta|\alpha)]}{\rm d}\alpha,\nonumber\\
    &=
    %\mathcal{K}_0(\beta)
    %\ln|\lambda_0^3(\beta)|/\beta
    %+
    \min_{\alpha\in\mathbb{R}}
    \left(
    \mathcal{F}^L_{\w}(\beta|\alpha)
    -\mathcal{S}_0(\alpha)/\beta\right),
    %+\mathcal{F}_0(\beta,\alpha),
    \label{F_final}
\end{align}
The free energy can also be written as the joint minimization, 
equivalent to a Legendre–Fenchel transformation when 
$\mathcal{S}_0(\alpha)$ is linear in $\alpha$,
\begin{equation}
    \mathcal{F}^L_{\w}(\beta)
    =
    %\mathcal{K}_0(\beta)
    %\ln|\lambda_0^3(\beta)|/\beta
    %+
    \min_{\alpha,\tD}\left(
    \w^\top\tD
    -[\mathcal{S}({\tD}|\alpha)+\mathcal{S}_0(\alpha)]/\beta\right).
    %+\mathcal{F}_0(\beta,\alpha).
    \label{F_final_2}
\end{equation}
Equation (\ref{F_final_2}) is our main result, an integration-free expression for 
the free energy for generalized linear MLIPs (\ref{linearE}).
The minimization over $\alpha$ and $\tD$ requires
\begin{equation}
    \nabla_\tD\mathcal{S}({\tD}|\alpha)=\beta\w,\quad
    \partial_\alpha\mathcal{F}^L_{\w}(\beta|\alpha)=
    \partial_\alpha\mathcal{S}_0(\alpha)
    \label{min_cond}
\end{equation}
which emphasizes the Legendre duality between $\beta\w$ and $\tD$.
Use of a harmonic reference energy (\ref{isosurface-H}) gives 
$\mathcal{S}_0(\alpha)=S_0+3\alpha/2$, equation (\ref{lim_dos}), 
as detailed in appendix \ref{app:isosurface_harmonic}, 
meaning $\partial_\alpha\mathcal{S}_0(\alpha)=3/2$.
Appendix \ref{app:isosurface_harmonic} also recovers familiar results for harmonic models; 
equation (\ref{min_cond}) reduces to the equipartition relation $\beta\langle E_0\rangle=3/2$.
\subsection{Gradients of the free energy}\label{sec:gradients}
The definition of the free energy (\ref{F_final}) as a double minimization over $\alpha$ and ${\tD}$
significantly simplifies the gradient of the free energy with respect to potential parameters $\w$ or thermodynamic variables such as temperature $1/\beta$. In particular, access to the 
gradient with respect to $\w$ allows the inclusion of finite temperature properties 
in objective functions for inverse design goals, a feature we 
explore in the numerical experiments. \\

With minimizing values $\alpha^*_{\beta,\w}$, $\tD^*_{\beta,\w,\alpha^*}$ 
for the isosurface and global descriptor, the $\w$-gradient is simply
\begin{align}
    \nabla_\w\mathcal{F}^L_{\w}(\beta)
    &=
    \tD^*_{\beta,\w,\alpha^*}\in\mathbb{R}^D.
    \label{gradient}
\end{align}
The internal energy is also a simple expression involving the minimizing vector 
$\tD^*_{\beta,\w,\alpha^*}$; when using $\hat{\alpha}(\X)$, defined in 
equation (\ref{isosurface-H}), we have
\begin{align}
    \mathcal{U}^L_{\w}(\beta)
    &=
    \partial_\beta(\beta\mathcal{F}^L_{\w}(\beta))
    =
    \frac{3}{2\beta}+\w^\top\tD^*_{\beta,\w,\alpha^*},
\end{align}
while when using $\hat{\alpha}(\X,\mom)$, defined in equation (\ref{isosurface-P}), 
the internal energy is simply $\mathcal{U}^L_{\w}(\beta)=\w^\top\tD^*_{\beta,\w,\alpha^*}$.
Evaluation of higher order gradients requires implicit 
derivatives\cite{blondel2022,maliyov2024exploringparameterdependenceatomic}, e.g. $\partial_\alpha \tD^*_{\beta,\w,\alpha^*}\in\mathbb{R}^D$, $\partial_\beta\alpha^*\in\mathbb{R}$ or $\partial_\w[\tD^*_{\beta,\w,\alpha^*}]^\top\in\mathbb{R}^{D\times D}$.

When using $\hat{\alpha}(\X)$ the specific heat at constant volume writes, 
in units of ${\rm k_B}$,
\begin{equation}
    {C^L_{V}}=3+
    \beta^2\left[\partial_\beta\tD^*_{\beta,\w,\alpha^*}
    +\partial_\beta\alpha^*_{\beta,\w}\partial_\alpha\tD^*_{\beta,\w,\alpha^*}\right]\w.
\end{equation}
Further exploration of finite temperature properties through thermodynamic relations, such as thermal expansion, will be the focus of future work. 
\subsection{Connection to free energy perturbation}\label{sec:SM-FEP}
The conditional free energy $\mathcal{F}_{\w}^L(\beta|\alpha)$ can be given by a
cumulant expansion, using $\langle\dots\rangle_\alpha$ for isosurface averages
\begin{equation}
    \mathcal{F}^L_{\w}(\beta|\alpha)
    =
    \langle\w^\top\tD\rangle_\alpha
    +
    \frac{N\beta}{2}\langle(\w^\top\delta\tD)^2\rangle_\alpha
    +\dots
    \label{cumulant_expansion}
\end{equation}
where $\delta\tD=\tD-\langle\tD\rangle$. The factor of $N$ to ensures intensivity 
of the covariance, as discussed in appendix \ref{app:intensive}.  
Free energy perturbation (FEP)\cite{zwanzig_fep_1954, rousset2010free,
castellano2024machinelearningassistedcanonical}, 
equation (\ref{cumulant_expansion_FEP}) also expresses the free energy difference
as a cumulant expansion over canonical averages with $E_0(\X)$. 
%\mcm{Now is nickel. I put the historical refs and book refs for FEP. I was completely puzzeled by MLACS refs ... }
As discussed in \ref{sec:gen_alpha}, as $N\to\infty$ canonical sampling at $\beta$
is equivalent to isosurface sampling at $\alpha=\alpha_\beta$, where the relation between 
$\beta$ and $\alpha_\beta$ depends on the form of the isosurface function $\hat{\alpha}(\X)$ or $\hat{\alpha}(\X,\mom)$, 
e.g. (\ref{isosurface-H}) or (\ref{isosurface-P}). 
%\mcmc{Now the discussion on FEP is very clear. Orders of magnitude better.} 
The FEP estimate is thus equivalent to a D-DOS estimate where we \textit{fix} 
${\alpha}=\alpha_\beta$, instead of minimizing over $\alpha$ as in equation (\ref{F_final}). 
When the free energy difference is very small, i.e. the target is
very similar to the reference, $\hat{\alpha}=\alpha_\beta$ may be a good approximate minimization. 
However, in the general case it is clear the D-DOS estimate can strongly differ from FEP estimates. 
This is evidenced later in Figures \ref{fig:BCC}b) and \ref{fig:BCC}c), 
where the minimizing $\alpha$ value at constant $\beta$ varies strongly with $\w$,
even at relatively low homologous temperatures (1000 K in W, around 1/4 of the melting temperature), 
while FEP would predict $\alpha$ to be constant with $\beta$.
\subsection{Errors in estimation via Laplace's method}\label{sec:laplace-error}
Estimation of free energies via (\ref{F_final}) clearly relies on our ability
to accurately approximate the conditional descriptor entropy function 
$\mathcal{S}({\tD}|\alpha)$ by some estimator, which here will be achieved 
by score matching in section \ref{sec:SM}. In this context, the curvature of 
$\mathcal{S}({\tD}|\alpha)$ in $\tD$ and $\alpha$ is crucial, both for 
the statistical efficiency of score matching and applicability of 
Laplace's method (appendix \ref{app:laplace}). In general, 
as might be expected, the prediction accuracy and statistical efficiency 
of any estimator will improve as the curvature increases in magnitude.
In close connection with existing free energy estimation schemes, 
selection of the reference energy function $E_0(\X)$ used to build 
$\hat{\alpha}(\X)$ or $\hat{\alpha}(\X,\mom)$ has a strong influence on 
the curvature of $\mathcal{S}({\tD}|\alpha)$ and thus any estimate 
$\mathcal{S}_\T({\tD|\alpha})$. A poorly chosen reference function 
will result in weaker curvature, as the descriptor distribution will vary less between 
isosurfaces $\hat{\alpha}(\X)$, imposing more stringent requirements on score matching estimates. 
These principles are evidenced in section \ref{sec:influence-score}, where we 
show use of a momentum-dependent isosurface function (\ref{isosurface-P}) results 
in larger curvature with $\alpha$ and lower errors in free energy estimates.
These considerations imply that a learnable isosurface function 
$\hat{\alpha}_{\boldsymbol{\phi}}(\X)$ should choose parameters 
$\boldsymbol{\phi}$ to maximize the curvature of the conditional 
entropy $\mathcal{S}_{\T;\boldsymbol{\phi}}({\tD}|\alpha)$, 
a direction we will explore in future work.\\ 

To summarize, the accuracy of Laplace's method for free energy estimation, 
equation (\ref{laplace}), will depend on our ability to 
determine the true minimum of $\mathcal{S}({\tD}|\alpha)$
from some noisy estimate $\mathcal{S}_\T({\tD|\alpha})$, which 
is strongly influenced by the choice of isosurface function $\hat{\alpha}$. 
The next section details the score matching procedure to produce such estimates.
\section{Score matching the conditional density of states\label{sec:SM}}
Evaluation of the free energy $\mathcal{F}_{\w}(\beta)$ in (\ref{F_final}) 
requires a minimization over the intensive descriptor entropy 
$\mathcal{S}(\tD|\alpha)$, equation (\ref{descriptor-entropy}). 
We will determine the parameters $\T$ of a model $\mathcal{S}_\T(\tD|\alpha)$ through score matching\cite{hyvarinen2005estimation}, to give estimates $\mathcal{F}_{\w;\T}(\beta)$ with score 
$\boldsymbol{\nabla}\mathcal{S}_\T(\tD|\alpha)\in\mathbb{R}^D$.\\

In appendix \ref{app:score_matching} we show the score matching loss\cite{hyvarinen2005estimation} 
reads, using $\langle\dots\rangle_\alpha$ for averages on isosurfaces $\hat{\alpha}(\X)=\alpha$,  
\begin{equation}
    \mathcal{L}(\T|\alpha)
    \equiv
    \left\langle\frac{N}{2}\|\boldsymbol{\nabla}\mathcal{S}_\T(\tD|\alpha)\|^2
    +
    \boldsymbol{\nabla}^2\mathcal{S}_\T(\tD|\alpha)\right\rangle_\alpha.
    \label{smloss}
\end{equation}
The factor of $N$ emerges from application of integration by 
parts\cite{hyvarinen2005estimation} in the derivation of (\ref{smloss}). 
Although $\mathcal{S}_\T(\tD|\alpha)$ is intensive, averages over $\alpha$ 
of $\mathcal{S}(\tD|\alpha)$ and it's gradients will give rise to terms $\mathcal{O}(N^{-s})$,
$s\geq0$ in $\mathcal{L}(\T|\alpha)$, giving a multiscale hierarchy\cite{pavliotis2008multiscale} 
$\mathcal{L}(\T|\alpha)=\sum_sN^{1-s}\mathcal{L}_s(\T|\alpha)$. As we detail in 
appendix \ref{app:score_matching}, each $\mathcal{L}_s(\T|\alpha)$ should 
be minimized recursively, ensuring the solution at $s=S$ 
does not affect the solution at $s<S$. In practice, this gave negligable 
improvement over simply minimizing (\ref{smloss}) directly, as for sufficiently 
large $N$ ($\mathcal{O}(100)$ in our experiments), a direct solution 
will naturally respect the dominant terms in the hierarchy, i.e. those for $s=0,1$.
\subsection{Low-rank compressed score models}
While the developments of section \ref{sec:Z-laplace} allow us 
to avoid high-dimensional integration, we still require a 
low-rank model to efficiently estimate and store any score model. In addition, 
the model should allow efficient minimization for free energy estimation 
via (\ref{F_final}). We use a common tensor compression 
approach\cite{sherman2020estimating} to produce a 
low-rank model for estimation of higher order moments.\\

Using $\langle\dots\rangle_\alpha$ to denote 
isosurface averages, we first {estimate} the 
isosurface mean $\hat{\bmu}_\alpha=\langle\tD\rangle_\alpha$ and 
intensive covariance $\hat{\bSig}_\alpha=N\langle\delta\tD\delta\tD^\top\rangle_\alpha$, 
where $\delta\tD=\tD-\bmu_\alpha$, a symmetric matrix which has $D$ 
orthonormal eigenvectors ${\bf v}_{\alpha,l}$, $l\in[1,D]$.
Our low-rank score model uses $F$ scalar functions 
${\bf f}(x)=[f_1(x),\dots,f_F(x)]\in\mathbb{R}^F$, 
with derivatives $\partial^n{\bf f}(x)
=[\partial^nf_1(x),\dots,\partial^nf_F(x)]\in\mathbb{R}^F$. 
We define the $D$ feature vectors of rank $F$:
\begin{equation}
{\bf f}_l(\tD|\alpha) \equiv {\bf f}([\tD-\bmu_\alpha]\cdot{\bf v}_{\alpha,l})
\in\mathbb{R}^{F},
\quad l\in[1,D].
\label{features}
\end{equation}
In practice, we use polynomial features of typical order $F=3-7$; 
we note that quadratic models ($F=2$) are insufficient to capture 
the anharmonic behavior of the D-DOS shown in section \ref{sec:experiments}. 
The conditional entropy model then reads, with ${\T}_l(\alpha)\in\mathbb{R}^F$,
\begin{equation}
    \mathcal{S}_\T(\tD|\alpha) \equiv
    \Theta_0(\alpha)+
    \sum_{l=1}^{D}
    {\bf f}_l(\tD|\alpha)
    \cdot{\T}_l(\alpha).
    \label{modelS}
\end{equation}
The conditional descriptor score then reads 
\begin{equation}
    \boldsymbol{\nabla}\mathcal{S}_\T(\tD|\alpha)
    = \sum_l 
    \left(
    \partial{\bf f}_l(\tD|\alpha)
    \cdot{\T}_l(\alpha)\right)
    {\bf v}_{\alpha,l}\in\mathbb{R}^{F},
\end{equation}
giving a score matching loss that is quadratic in ${\T}_l(\alpha)$;
by the orthonormality of the ${\bf v}_{\alpha,l}$, 
minimization reduces to solving the $D$ linear equations of rank $F$:
\begin{equation}
    N\langle
    \partial{\bf f}_l(\tD|\alpha)
    [\partial{\bf f}_l(\tD|\alpha)]^\top
    \rangle_\alpha
    {\T}_l(\alpha)
    =-\langle\partial^2{\bf f}_l(\tD|\alpha)\rangle_\alpha.
    \label{linearSM}
\end{equation}
Solution of (\ref{linearSM}) for each $\alpha$ fixes ${\T}_l(\alpha)$,
while the constant $\Theta_0(\alpha)\in\mathbb{R}$ is determined by 
equation (\ref{szero}), i.e. ensures $\max_\tD\mathcal{S}_\T(\tD|\alpha)=0$.
For some model $\w$, the conditional free energy (\ref{condF_final}) then reads
\begin{equation}
    \mathcal{F}^L_{\w;\T}(\beta|\alpha)\equiv
    \min_{\tD} \left(\w\cdot\tD-\mathcal{S}_\T(\tD|\alpha)/\beta\right),
\end{equation}
%\mcmc{It is not better to use parenthesis after $\min \left( \right)$ ? Not only here ...  } 
which is achieved when $\nabla_\tD\mathcal{S}_\T(\tD|\alpha)=\beta\w$. 
We can then interpolate $\mathcal{F}^L_{\w;\T}(\beta|\alpha)$ the 
sampled range of $\alpha$ values to give a final free energy estimate 
of
\begin{equation}
    \mathcal{F}^L_{\w;\T}(\beta)\equiv
    \min_\alpha \left(\mathcal{F}^L_{\w;\T}(\beta|\alpha) - \mathcal{S}_0(\alpha)/\beta\right),
    \label{modelF}
\end{equation}
which is achieved when $\partial_\alpha\mathcal{S}_0(\alpha)=\beta
\partial_\alpha\mathcal{F}^L_{\w;\T}(\beta|\alpha)$. The final minimizing 
values of the descriptor vector $\tD^*$ allows evaluation of the gradient
$\partial_\w \mathcal{F}^L_{\w;\T}(\beta)=\tD^*$, equation (\ref{gradient}).
Equation (\ref{modelF}) is the central result of this paper, a closed-form expression for the 
vibrational free energy of linear MLIPs (\ref{linearE}). 
Section \ref{sec:numerics} details numerical implementation of 
the score matching procedure and section \ref{sec:experiments} presents 
verification of predictions (\ref{modelF}) against calculation of 
$\mathcal{F}^L_{\w}(\beta)$ using thermodynamic integration, equation (\ref{TI}).

\subsection{Error analysis and prediction}\label{sec:error_analysis}
To estimate errors on the free energy $\mathcal{F}^L_{\w;\T}(\beta)$, 
we can use standard error estimates to determine the uncertainty
on the isosurface mean $\bmu_\alpha$ and covariance eigenvectors ${\bf v}_{\alpha,l}$
to produce errors $\delta{\bf f}_l(\tD)$ on feature vectors (\ref{features}). 
In addition, epistemic uncertainties on expectations 
in the score matching loss (\ref{linearSM})
will give uncertainties $\delta{\T}_l(\alpha)$ on model coefficients 
${\T}_l(\alpha)$, which can
be estimated by either subsampling the simulation data to produce an ensemble 
of model coefficients or extracting posterior uncertainties from Bayesian 
regression schemes\cite{von2011bayesian}. Propagating these combined 
uncertainties provides a reasonable and efficient estimate of 
sampling errors, as we demonstrate in section \ref{sec:experiments}.\\

While these estimates are of comparable 
accuracy to available error estimates for thermodynamic sampling\cite{castellano2024machinelearningassistedcanonical}, 
a true error estimate should account for the misspecification
of any low-rank score matching model, which in general requires study 
of the generalization error\cite{masegosa2020} rather than the score matching loss. 
Indeed, we use our recently introduced misspecification-aware regression
scheme\cite{swinburne2025} to determine uncertainty in MLIP parameters $\w$
when fitting against \textit{ab initio} reference data. As discussed in \ref{sec:conclusions} 
work will focus on developing such a scheme in tandem with learnable isosurface 
functions (section \ref{sec:laplace-error}) to provide rigorous guarantees 
on score matching estimates. 

\subsection{Systematic error correction}
As we detail in section \ref{sec:experiments}, we find 
the estimated D-DOS errors to be excellent predictors of the 
observed errors. In addition, both predicted and observed errors 
are typically very low, around 1-2 meV/atom, rising to 10 meV/atom if the reference 
model is poorly chosen or the system is particularly anharmonic. 
As we show in section \ref{sec:influence-score}, these errors can be 
largely corrected through the use of a momentum-dependent isosurface
(\ref{isosurface-P}) bringing observed and predicted errors back
within the stringent 1-2 meV/atom threshold.\\

However, if tightly converged ($<$1meV/atom) estimates of the free energy 
are desired for a given parameter choice $\w$, the close connection between
the D-DOS conditional free energy $\mathcal{F}^L_{\w}(\beta|\alpha)$
and free energy perturbation (FEP), discussed in section \ref{sec:SM-FEP},
offers a systematic correction scheme. 
Any predicted value of $\mathcal{F}^L_{\w;\T}(\beta|\alpha)$ from our 
score matching estimate can be updated through short isosurface sampling runs,
recording the \textit{difference} between observed cumulants of $\w^\top\tD$
and those predicted by $\mathcal{S}_{\w;\T}(\tD|\alpha))$, conducted over 
a small range of $\alpha$ to account for updated moments changing 
the minimum solution
$\partial_\alpha\mathcal{S}_0(\alpha)=\beta\partial_\alpha\mathcal{F}_w(\beta|\alpha)$.
Following established FEP techniques\cite{grabowski2019ab,castellano2024machinelearningassistedcanonical}
this procedure can then be extended to include \textit{ab initio} data. 
However, given the accuracy of our D-DOS estimations in section \ref{sec:experiments}, 
we focus on exploring the unique abilities of the D-DOS scheme in forward and back parameter 
propagation, leaving a study of this correction scheme to future work.

\section{Numerical implementation}\label{sec:numerics}
In this section, we describe in detail how the D-DOS sampling scheme 
is implemented, and how a broad ensemble of free energy 
estimates was produced using thermodynamic sampling in order to 
provide stringent tests of D-DOS free energy estimates. 
We describe the low-rank linear MLIPs employed (\ref{sec:snap}), 
the production of DFT training data (\ref{sec:dft}) 
the production of reference free energy estimates via 
thermodynamic integration (\ref{TI}) and details of 
the D-DOS score matching campaign (\ref{sec:ddos-sampling}).
We focus on MLIPs that approximate the bcc, A15 and fcc phases 
of tungsten (W), molybdenum (Mo) and iron (Fe)\cite{goryaeva2021efficient}.

\subsection{Choice of linear MLIP}\label{sec:snap} 
We build a linear MLIP using the bispectral BSO(4) descriptor functions, first introduced in the 
\texttt{SNAP} MLIP family\cite{Thompson_snap_2015}. While a quadratic 
featurization is often used\cite{goryaeva2021,grigorev2023calculation} 
we employ the original linear model, i.e. $\tD_i = \hat{\bphi}(\D_i) = \D_i  \in \mathbb{R}^d$. 
For unary systems we have $H=4$ hyperparameters ${\bf h}$, the cutoff radius 
$r_c$, the number of bispectrum components $D$ and two additional weights in 
the representation of the atomic density. We refer the reader to the original 
publications for further details\cite{Thompson_snap_2015}. To test the 
transferability of the sampling scheme under different reference models 
$E_0(\X)$, we fix $\bf h$ to be the same for all potentials, regardless of the 
specie in training data, using a cutoff radius of $r_c=4.7${\AA}  
and $D=55$ bispectrum components. While we consider models approximating Mo and W 
(see section \ref{sec:dft}), which have similar equilibrium volumes, 
we note that the bispectrum descriptor is invariant\cite{Thompson_snap_2015} 
under a homogeneous rescaling of both the atomic configuration and the cutoff radius, 
i.e. the CD-DOS is invariant for fixed $V/r^3_c$.\\

We expect the numerical results of this section to hold directly if 
we replace the bispectrum descriptor with other "low-dimensional" 
($d=D=\mathcal{O}(100)$) descriptors such as \texttt{POD}\cite{Nguyen2023} or 
hybrid descriptors in \texttt{MILADY}\cite{goryaeva2019towards, dezaphie_designing_2025}. For models such as
\texttt{MTP}\cite{Shapeev_MTP} or \texttt{ACE}\cite{lysogorskiy2021performant}, where $d=\mathcal{O}(10^3)$, the features or score model (or both) will accept some rank reduction, e.g. a linear projection $\tD_i(\D_i)={\bf P}\D_i$, where ${\bf P}\in\mathbb{R}^{D\times d}$, with $D=\mathcal{O}(100)$. 
The \texttt{POD} scheme applies rank-reduction to the 
radial part of the descriptor, following \cite{goscinski_optimal_radial_2021}.
Many other rank-reduction schemes have been proposed in recent years, including 
linear embedding\cite{darby_compressing_soap_2022, willatt_atom-density_2019} 
or tensor sketching \cite{darby2023tensor}.

\subsection{Training data for Fe, W and Mo}\label{sec:dft}

The majority of the database configurations for Fe and W are those published in \cite{goryaeva2021}. The W database originates from the defect- and dislocation-oriented database in \cite{goryaeva2021}, which was modified and updated with molecular dynamics instances in \cite{Zhong2023} to improve its suitability for finite-temperature calculations and thermoelasticity of W. 
Finally, for this study, using the MLIP developed in \cite{Zhong2023}, we prepared multiple samples of W in the A15 phase or liquid within the NPT ensemble, covering temperatures from 100 K to 5000 K. Each system contained 216 atoms. We selected 96 snapshots, which were then recomputed using the same DFT parameterization as in \cite{goryaeva2021, Zhong2023}.
The Mo database was specifically designed for this study to ensure a well-represented configuration of Mo at high temperatures in the bcc and A15 phases. The detailed components of the database, as well as the ab initio details, are described in the Appendix \ref{db:mo}. 

\subsection{Ensemble of potential parameters for testing in forward-propagation}\label{subsec:pot_params}
From the DFT training databases for $x=\rm W,Mo,Fe$ (\ref{sec:dft}),
we generate a broad range of parameter values $\w\in\W_x$ for \texttt{SNAP} MLIPs (\ref{sec:snap})
using a recently introduced \cite{swinburne2025} Bayesian linear regression scheme. 
The scheme is designed to produce robust parameter uncertainties for misspecified surrogate models 
of low-noise calculations, which is precisely the regime encountered when fitting linear MLIPs to DFT data. 

\begin{figure}[!t]
    \includegraphics[width=0.99\linewidth]{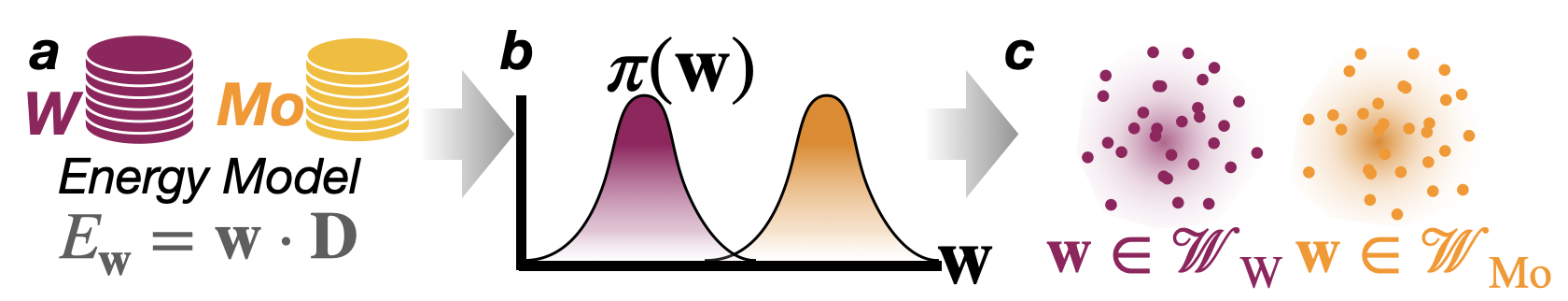}
    \caption{Producing ensemble of potential parameters for testing D-DOS in forward propagation.
    \textbf{a}) A linear MLIP (\ref{linearE}) form is chosen to approximate \textit{ab initio} databases, here of Mo and W. 
    \textbf{b}) Misspecification-aware Bayesian regression\cite{swinburne2025} returns  
    a parameter posterior which is broad for simple MLIPs and diverse databases. 
    \textbf{c}) The posterior is sampled to produce an ensemble of stable 
    model parameters $\w\in\mathcal{W}_{\rm W}$ and $\w\in\mathcal{W}_{\rm Mo}$ 
    used for testing.\label{fig:ensemble}}
\end{figure}

Taking training data for $x=\rm W$ or $\rm Mo$, the method produces a posterior 
distribution $\pi(\w)$ (Figure \ref{fig:ensemble}a-b), with strong guarantees that posterior 
predictions bound the true DFT result, irrespective of how each training point is weighted.
As the \texttt{SNAP} form has a relatively small number ($\mathcal{O}(100)$) of adjustable parameters 
it is strongly misspecified (large model-form error) to the diverse training database and 
thus the posterior distribution gives a broad range of parameter values. 

Each training point was weighted using a procedure described elsewhere \cite{goryaeva2021, Zhong2023}.
While we also explored randomly varying weights associated with defects and other disordered structures, 
in all cases we maintained consistently high weights for structures corresponding to small deformations of the cubic unit cell in the bcc, fcc, or A15 phases. This procedure ensures that the resulting potential ensemble 
yields lattice parameters within a range of \(10^{-4}\) \AA\ and elastic constants that follow a narrow distribution centered around the target DFT average values. %

We construct our ensemble $\W_x$, $x=\rm W,Mo,Fe$ by applying CUR sparsification \cite{drineas_fast_2006, drineas_cur_relative-error_2008} to a large set of posterior samples to extract \(\mathcal{O}(100)\) parameter
vectors which show sufficient dynamical stability to  allow for convergence when performing thermodynamic integration at high temperature (Figure \ref{fig:ensemble}c). We also identify a `reference' value $\bar{\w}_x$, being a stable parameter choice that has the optimal error to training data, i.e. the best overall interatomic potential choice. For each parameter $\w$ we have computed the free energy $\mathcal{F}^L_\w(\beta)$, as is detailed in the section \ref{sec:TI-sampling}. 

\subsection{Free energies from thermodynamic integration}\label{sec:TI-sampling}

With a given choice of MLIP parameters $\w$, we employ a recently introduced
thermodynamic integration method\cite{cao2014,Zhong2023} to calculate the corresponding 
NVT phase free energies $\mathcal{F}^L_\w(\beta)$, 
equation (\ref{limit_free_energy}). The thermodynamic scheme first 
calculates the Hessian matrix ${\bf H}_\w$ for a given parameter choice, 
to give a harmonic free energy prediction and to parametrize a 'representative' harmonic reference. 
Rather than the sequential integration over $\eta$ as described by equation (\ref{TI}), 
the employed scheme instead uses a Bayesian reformulation to sample all $\eta\in[0,1]$ values simultaneously, which significantly accelerates convergence\cite{cao2014}. In addition, a `blocking' constraint is used to prevent trajectories escaping the metastable basin of any crystalline phase. 
We refer the reader to\cite{Zhong2023} for further details.

Even with these blocking constraints, in many cases phases had poor metastability at high temperatures, in particular the A15 phase, 
which was rectified by adding more high temperature A15 configurations to training data 
and restricting the range of potential parameters. These dynamical instabilities 
reflect general trends observed in long molecular dynamics trajectories, 
where high-dimensional MLIPs are prone to failure over long time simulations\cite{ibayashi_allegro-legato_2023, vita_data_2023}. 

While there is currently no general solution to the MLIP stability problem, 
even for the relatively low-dimensional ($D=\mathcal{O}(100)$) descriptors
used in this study, it can be mitigated by enriching the training database\cite{Zhong2023}. 
In contrast, our score matching procedure only requires stability of the Hessian matrix ${\bf H}$ 
used for the harmonic reference potential $E_0(\X)$, equation (\ref{isosurface-H}), 
a much weaker condition than dynamical stability. The observed accuracy, detailed below, 
strongly suggests our sampling scheme may be able to predict phase free energies for a much 
broader range of parameter space than those that can be efficiently sampled via traditional methods.
A full exploration of this ability is one of the many future directions we discuss in the conclusions 
(\ref{sec:conclusions}).

The final sampling campaign to generate reference free energies for comparison against D-DOS estimates 
required around $\mathcal{O}(10^4)$ CPU hours, or $\mathcal{O}(10^5)$ force calls per model, with 
blocking analysis\cite{rousset2010free,Athenes2017} applied to estimate the standard error in each free energy estimate. We emphasize that the scheme described in this section represents the state-of-the-art 
in free energy estimation for MLIPs. Nevertheless, for any given choice of model parameters,
free energy estimation requires at least $\mathcal{O}(10^6)$ CPU hours, 
irrespective of available resources, which significantly complicates uncertainty quantification 
via forward propagation and completely precludes including finite temperature
properties during model training via back-propagation. The model-agnostic 
D-DOS scheme detailed introduced in this paper provides a first general solution
for MLIPs that can be cast into the general linear form (\ref{linearE}).

\subsection{D-DOS score matching sampling campaign}\label{sec:ddos-sampling}
As detailed in section \ref{sec:SM}, when using the harmonic isosurface 
function (\ref{isosurface-H}), our score matching sampling campaign 
reduces to sampling descriptor distributions on isosurfaces $\hat{\alpha}(\X)=\alpha$ 
defined by the Hessian $\bf H$ of some reference potential $E_0(\X)$. For 
momentum-dependent isosurface functions (\ref{isosurface-P}) we instead record 
samples from an ensemble of short NVE runs, which we explore in section \ref{sec:influence-score}.
We tested the harmonic isosurface function (\ref{isosurface-H}) using one
Hessian ${\bf H}={\bf H}_{x}$ per phase for $x=\rm W, \, Mo,\ Fe$. Each Hessian was 
calculated using the appropriate lattice structure and the reference (loss minimizing)
potential parameters $\w=\bar{\w}_x$ described in the previous section.
With a given isosurface function $\hat{\alpha}(\X)$, we generated $\mathcal{O}(10^3)$ 
independent samples on $\hat{\alpha}(\X)=\alpha$ for a range of $\alpha$ values at constant volume. 
It is simple to distribute sampling across multiple processors, as the harmonic isosurface samples 
are trivially independent (see appendix \ref{app:isosurface_harmonic}).
This enables a significant reduction in the wall-clock time for 
sampling over trajectory-based methods such as thermodynamic integration.\\
Our open-source implementation\cite{descriptordos} uses \texttt{LAMMPS}\cite{LAMMPS} 
to evaluate \texttt{SNAP}\cite{Thompson_snap_2015} descriptors; as 
a rough guide, with $N=\mathcal{O}(10^2)$ atoms, the sampling campaign used to 
produce the results below required around $\mathcal{O}(10)$ 
seconds per $\alpha$ value on $\mathcal{O}(10^2)$ CPU cores.
A converged score model built from $\mathcal{O}(10)$ $\alpha$-values was thus 
achieved in under 5 minutes at each volume $V$ and Hessian choice ${\bf H}$. 
Algorithm \ref{D-DOS-algo} outlines the sampling campaign for each of 
the two isosurface functions we employ: $\hat{\alpha}(\X)$, equation (\ref{isosurface-H}), or $\hat{\alpha}(\X,\mom)$, equation (\ref{isosurface-P}).
Use of momentum-dependent isosurfaces $\hat{\alpha}(\X,\mom)$ 
requires a single free energy estimate, which could be either from 
a separate D-DOS estimation or `traditional` sampling methods. In addition, 
to allowing for NVE sample decorrelation gives a factor 10 greater sampling effort, 
i.e. comparable with the effort for a single fixed model sampling. 
Computational demands quoted are when using $\hat{\alpha}(\X)$; 
future work will investigate schemes to 
further accelerate momentum-dependent $\hat{\alpha}(\X,\mom)$.\\

The final score model requires minimal storage, being only the $\mathcal{O}(100)$ scalars contained 
in the vector $\T_\alpha$, equation (\ref{modelS}), over a range of $\alpha$ values at constant $V$, ${\bf H}$. 
It is therefore possible to efficiently store many score models to investigate the
influence of the reference model on free energy predictions. For example, section 
\ref{sec:alchem} demonstrates how a D-DOS using ${\bf H}$ from 
$\bar{\w}_{\rm Mo}\in\mathcal{W}_{\rm Mo}$ can predict free energies from the 
W ensemble, $\w\in\mathcal{W}_{\rm W}$.\\

Table \ref{tab:cost_comp} provides a rough guide to the computational cost of existing 
methods, as reported in recent works\cite{Zhong2023,menon2024electrons,castellano2024machinelearningassistedcanonical}, 
alongside the D-DOS sampling scheme detailed above. 
As can be seen, D-DOS is at least an order of magnitude more efficient than TI and up to two orders of magnitude more efficient than AS, even before considering the massive reduction in wall-clock time due to parallelization.
We again emphasize that in addition to the modest computational requirements 
of D-DOS, sampling is \textit{model-agnostic}, only performed for a given choice of descriptor hyperparameters, system volume and function $\hat{\alpha}(\X)$ used for isosurface construction. 
Model agnosticism is the key innovation of the D-DOS approach, allowing 
rapid forward propagation for uncertainty quantification and, uniquely, back-propagation
for inverse design goals. These unique abilities are demonstrated and tested in the 
next section. 

\begin{algorithm}[H]
    \caption{Sampling for NVT estimator $\mathcal{F}_{\w;\T}(\beta)$\label{D-DOS-algo}}
    \begin{algorithmic}
    \State Crystal phase $p\in\mathcal{P}$, atom count $N$ and cell volume $V$
    \State Reference potential $E_0(\X)$, typically $\bar{\w}^\top\tD$, $\bar{\w}\in\mathcal{W}$.
    \If{$\hat{\alpha}=\hat{\alpha}(\X)$}
    \State Calculate $\bf H$ of $E_0(\X)$ to give $\mathcal{F}_0(\beta)$ and thus $\mathcal{S}_0(\alpha)$
    \Else{\quad$\hat{\alpha}=\hat{\alpha}(\X,\mom)$}
    \State Calculate true $\mathcal{F}_0(\beta)$ and $\mathcal{U}_0(\beta)$ of $E_0(\X)$
    to give $\mathcal{S}_0(\alpha)$
    \EndIf
    \State Isosurface values $\alpha\in\mathcal{A}$ with $\alpha=\ln|\mathcal{U}_0(\beta)/{\rm U}_0|$
    \For{ $\alpha\in\mathcal{A}$}
    \For{$n\in N_{\rm samples}$, across parallel workers}
    \If{$\hat{\alpha}=\hat{\alpha}(\X)$}
    \State Generate $\delta\X$ such that $\delta{\X}{\bf H}\delta{\X}$=$2N{\rm U}_0\exp(\alpha)$
    \State Record $\tD$
    \Else{\quad$\hat{\alpha}=\hat{\alpha}(\X,\mom)$}
    \State Short NVE runs at $E={\rm U}_0\exp(\alpha)$
    \State Record $\tD\oplus K$
    \EndIf
    \EndFor
    \State Estimate $\bmu_\alpha$, $\bSig_\alpha$, solve (\ref{linearSM}) for vector $\T_\alpha$
    \EndFor
    \State Store $\T=\{\T_\alpha\}_{\alpha\in\mathcal{A}}$, NVT estimator $\mathcal{F}_{\w;\T}(\beta)$,  $\forall\w\in\mathcal{W}$.
    \end{algorithmic}
\end{algorithm}

\begin{table}[h!]
\centering
\begin{tabular}{|c|c|c|c|c|}
\hline
\textbf{Method} & \textbf{$|\Delta\mathcal{F}|$}  & \textbf{Steps} & 
\textbf{Steps/Worker} & \textbf{Agnostic}
\\ \hline
FEP\cite{Book_Frenkel} (\ref{sec:FEP}) & 10 & $\sim10^6$ & $\sim10^4$ & No
\\ 
TI\cite{Athenes2017,Zhong2023} (\ref{sec:TI})
& 150& $\sim10^6$ & $\sim10^5$ & No
\\ 
AS\cite{menon2024electrons} (\ref{sec:AS}) & 150  & $\sim10^8$ & $\sim10^7$ & No
\\ 
\textbf{D-DOS}
& \textbf{200} & $\bf\sim10^5$ & $\bf\sim10^2$ & \textbf{Yes}\\ 
\hline
\end{tabular}
\caption{Approximate comparison of computational cost for various methods 
to calculate solid-state vibrational free energies. 
$|\Delta\mathcal{F}|$: approximate maximum free energy difference that can be targeted at 1000 K in meV/atom, rising approximately proportionally with temperature.
Steps: approximate number of force evaluations for 1-2 meV/atom convergence. 
Steps/Worker: approximate number of force evaluations per worker in a parallel sampling 
scheme, indicating the minimum wall-time. Agnostic: D-DOS sampling 
only needs to be conducted \textit{once} for uniform approximation ability across broad 
range of potential parameters, while all other methods must be repeated, massively 
increasing the computational burden. 
\label{tab:cost_comp}}
\end{table}
\section{Numerical experiments}\label{sec:experiments}

In this section we detail numerical experiments, testing the 
D-DOS free energy estimates in forward and back-propagation. 
In forward propagation, our aim is to predict the free energy 
of all models $\w\in\mathcal{W}$ in a given ensemble, 
over a broad range of temperatures, from a single score matching campaign.\\

Accuracy in forward propagation, combined with robust misspecification-aware
parameter uncertainties\cite{swinburne2025},
clearly ensures accurate uncertainty quantification of free energies. 
A detailed investigation of D-DOS uncertainty quantification will be 
presented in a separate study, following recent work 
on static properties\cite{perez2025uncertaintyquantificationmisspecifiedmachine}.
In back-propagation, we exploit access to parameter gradients of the free energy, 
equation (\ref{gradient}), to fine-tune a given interatomic
potential to match finite temperature properties. To our
knowledge, this is the first demonstration of back-propagation
being used to target phase boundaries in atomic simulation.\\

Section \ref{sec:nvt-free} presents tests in forward propagation, 
predicting NVT free energies for bcc and A15 phases using a harmonic 
reference potential in $\hat{\alpha}(\X)$, equation (\ref{isosurface-H}). 
Section \ref{sec:influence-score} shows how predictions for the strongly 
anharmonic A15 phase can be systematically improved using the momentum-dependent 
isosurface function (\ref{isosurface-P}), invoking the curvature considerations 
raised in \ref{sec:laplace-error}. Section \ref{sec:alchem} demonstrates the excellent 
alchemical transferability of the D-DOS approach, taking a D-DOS estimator using a 
harmonic reference for Mo to predict NVT free energies of models for W. Finally, 
sections \ref{sec:nvt-target} and \ref{sec:alpha-gamma-tr} demonstrate 
how the D-DOS approach can be used in back-propagation for inverse design goals,
starting from some reference potential $\bar{\w}$. Section \ref{sec:nvt-target}
adjusts the potential to match a given set of $\mathcal{F}(\beta)$ observations, 
regularizing against the original fit. Section \ref{sec:alpha-gamma-tr} extends 
this principle to minimize the $\alpha-\gamma$ free energy difference at a 
desired target temperature, targeting a phase boundary. 

\subsection{Prediction of NVT free energies}\label{sec:nvt-free}
\begin{figure}[!t]
    \includegraphics[width=0.99\linewidth]{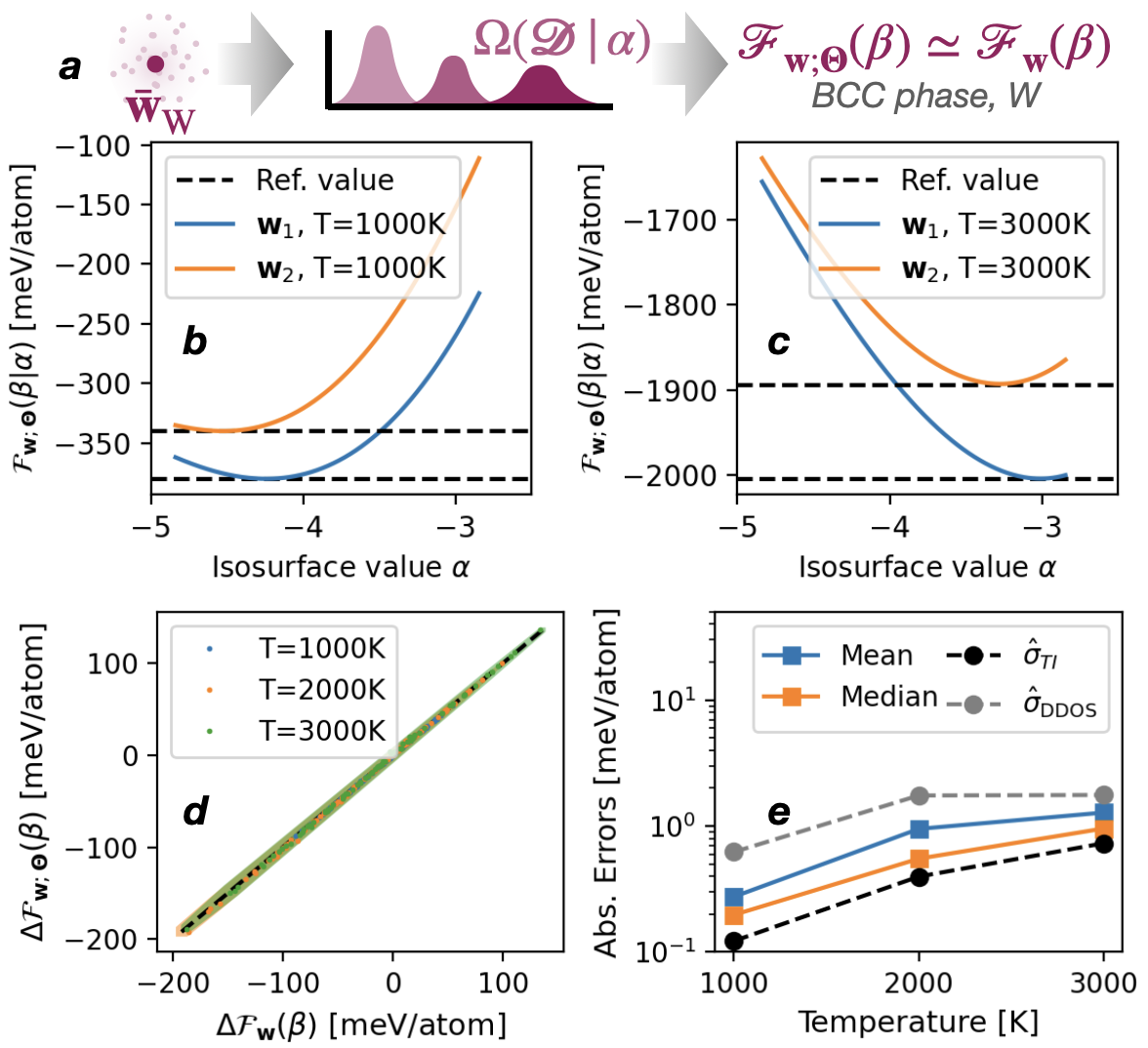}
    \caption{NVT free energy predictions in bcc phase of \texttt{SNAP} models $\w\in\W_{\rm W}$ 
    from tungsten database. \textbf{a)} D-DOS sampling used $\hat{\alpha}({\bf X})$ function 
    (\ref{isosurface-H}) using Hessian from $\bar{\w}_{\rm W}\in\W_{\rm W}$.
    \textbf{b-c)} Conditional free energy (\ref{condF_final}) $\mathcal{F}_{\w;\T}(\beta|\alpha)$
    with $\alpha$ at T=1000 K (left) and T=3000 K (right). 
    Horizontal dashed lines give reference values from thermodynamic integration.
    \textbf{d)} Parity plot for three temperatures. Note the large (300 meV) 
    spread of values at high temperatures. Shaded areas are convex hull around 
    D-DOS predictions with errors, i.e. $\mathcal{F}_{\w;\T}(\beta|\alpha)\pm\hat{\sigma}_{\rm DDOS}$.
    \textbf{e)} Mean and median absolute 
    errors, showing sub meV/atom accuracy, along with estimations from thermodynamic 
    integration ($\sigma_{\rm TI}$) and D-DOS ($\hat{\sigma}_{\rm DDOS}$).}\label{fig:BCC}
\end{figure}

Our first results demonstrate the ability of our D-DOS sampling scheme to predict 
NVT free energies $\mathcal{F}_\w(\beta)$ for the bcc and A15 phases, 
using the ensemble of potentials $\w\in\W_{x}$, $x=\rm W,Mo$ described above 
and illustrated in Figure \ref{fig:BCC}a).  
Figure \ref{fig:BCC} shows D-DOS predictions for bcc phases against 
the corresponding thermodynamic integration (TI) calculations. The D-DOS was estimated 
by score matching on isosurfaces $\hat{\alpha}(\X)=\alpha$, equation (\ref{isosurface-H}),
using a \textit{single} Hessian matrix ${\bf H}$ from the same ensemble. 
Panels b) and c) display how $\mathcal{F}_{\w;\T}(\beta|\alpha)$ 
depends on $\alpha$ for two different potentials $\w_1,\w_2\in\mathcal{W}_{\rm W}$.
The minimization procedure can be efficiently achieved, with the full free energy 
estimation procedure requiring around 10-100 microseconds of CPU effort depending on the 
complexity of the score model employed; in the current implementation, 
the effort scales approximately linearly with the number of feature functions $F$. 
Panels d) and e) show the excellent approximation ability of the D-DOS estimator despite 
the high degree of diversity across the ensemble. Importantly, the predicted 
D-DOS errors are excellent estimates of the actual errors. In \ref{fig:BCC}d) 
we subtract the free energy of the harmonic system used to build $\hat{\alpha}(\X)$,
thus displaying the explicit (NVT) anharmonicity captured by the D-DOS estimator, 
with typical absolute values of 150 meV/atom at 3000K, with ensemble variations 
of around 300 meV/atom. The mean absolute errors of 1 meV/atom, or 1/40 kcal/mol, 
even at these elevated temperatures, represent a key numerical result of this paper, 
showing that the presented model-agnostic approach is both more efficient and 
directly comparable to existing state-of-the-art sampling approaches. \\

\begin{figure}[!t]
    \includegraphics[width=0.99\linewidth]{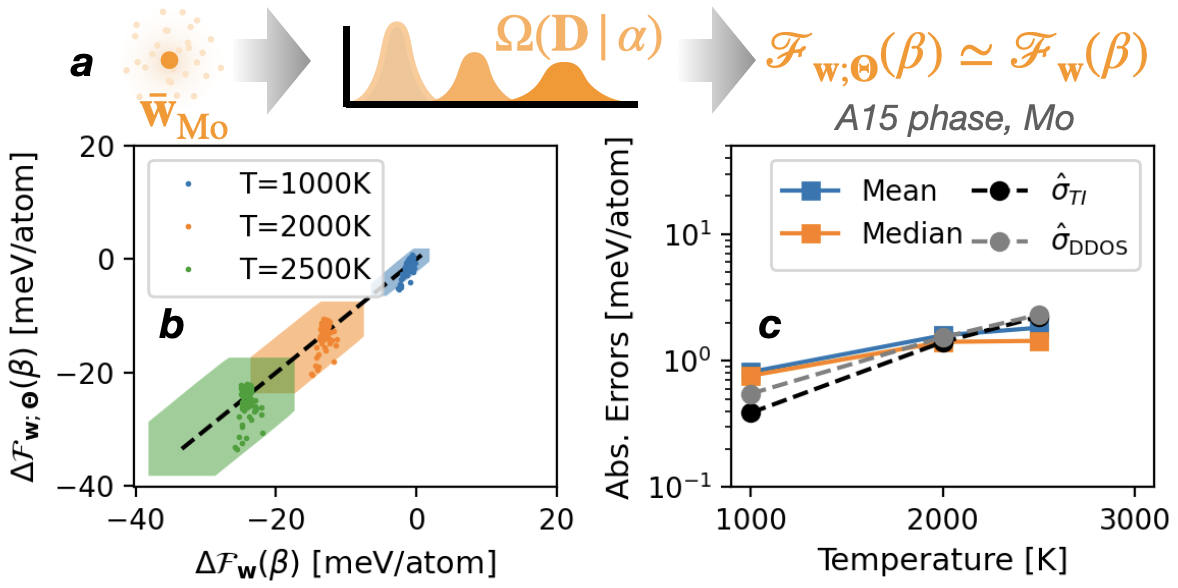}
     \caption{NVT free energy predictions in A15 phase of \texttt{SNAP} models 
     from molybdenum database. \textbf{a)} D-DOS sampling employed $\hat{\alpha}({\bf X})$,
     equation (\ref{isosurface-H}), built with $\bar{\w}_{\rm Mo}$.  
     The testing potential ensemble $\w\in\W_{\rm Mo}$ was less diverse to ensure stability of TI calculations. \textbf{b)} Parity plot at three temperatures. \textbf{c)} Mean and median absolute errors over $\mathcal{W}_{\rm Mo}$.
    }\label{fig:Mo-A15}
\end{figure}

Figure \ref{fig:Mo-A15} shows D-DOS predictions for the NVT free energy of the 
metastable A15 phases for parameters from the Mo ensemble, $\w\in\W_{\rm Mo}$.
The metastability of the A15 phase significantly reduces the diversity of potentials
whose free energy can stably be estimated via TI. Similarly, the 
estimated errors in the final values from TI are 
correspondingly larger. For D-DOS, however, there are no stability issues- we only require 
the potential $\bar{\w}\in\W_{\rm Mo}$ used to build the isosurface has a Hessian with 
no negative eigenvalues. This opens many perspectives for sampling of unstable phases, 
as we discuss in \ref{sec:conclusions}. 
As can be seen in figure \ref{fig:A15} the D-DOS estimates 
retain mean average errors of less than 2meV/atom at 2500 K, and, crucially, 
the propagated error estimates envelope the true errors. At higher temperatures, 
we see that the ensemble average sampling errors from D-DOS and TI are almost 
identical.

\subsection{Influence of the isosurface function}\label{sec:influence-score}
\begin{figure}[!t]
    \includegraphics[width=0.99\linewidth]{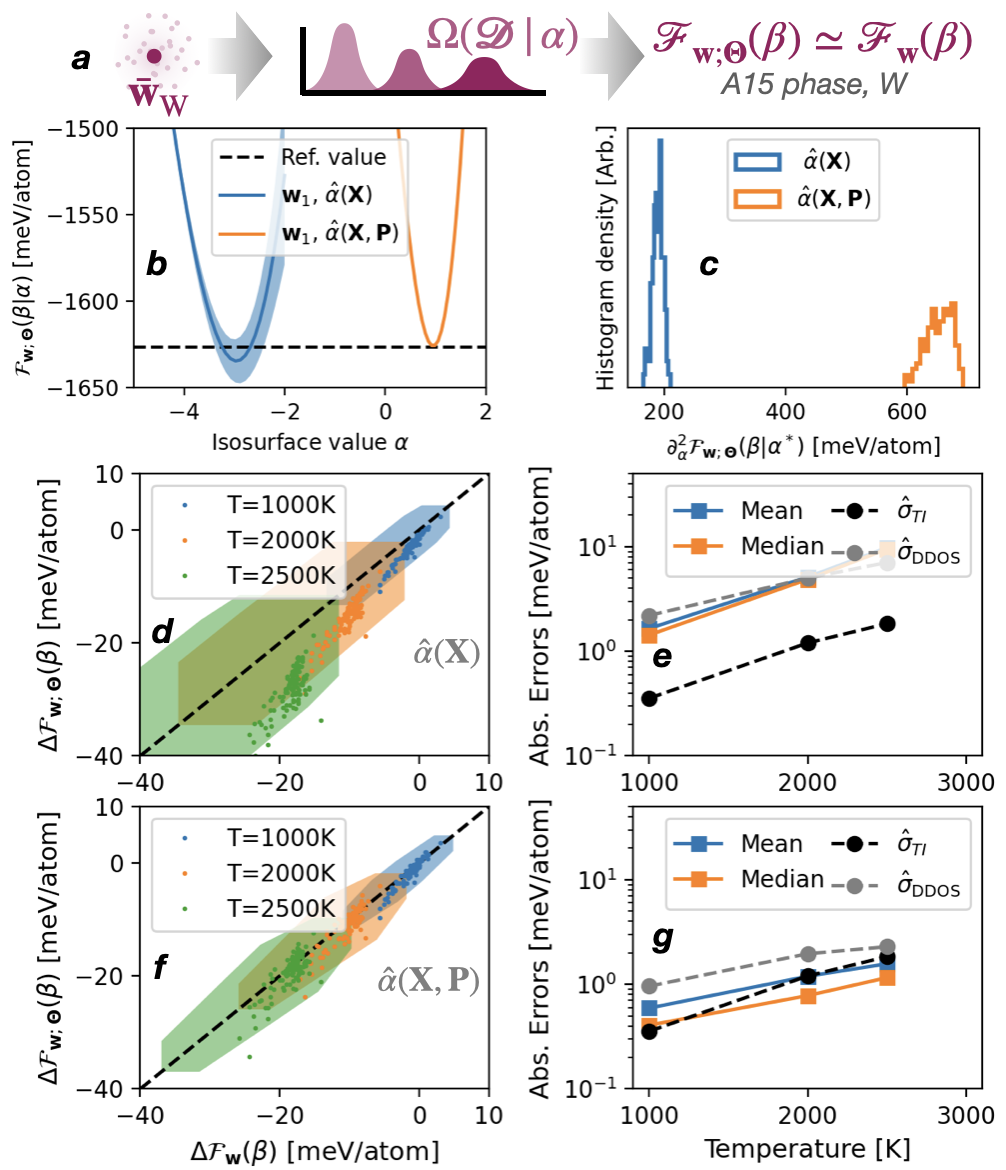}
    \caption{Influence of isosurface function in NVT free energy predictions of 
    \texttt{SNAP} models for W in the A15 phase, a strongly anharmonic structure 
    which was only stable in a tightly clustered set of ensemble members $\w\in\mathcal{W}$.
    \textbf{a)} D-DOS sampling employed
    $\hat{\alpha}(\X)$ and $\hat{\alpha}(\X,\mom)$ with $\w=\bar{\w}_{\rm W}$ (see \ref{sec:gen_alpha}).
    \textbf{b)} Conditional free energy (\ref{condF_final}) at T=2500 K, with propagated errors,
    for a representative potential $\w_1\in\mathcal{W}_{\rm W}$, using $\hat{\alpha}(\X)$ or $\hat{\alpha}(\X,\mom)$. 
    \textbf{c)} $\hat{\alpha}(\X,\mom)$ gives a conditional free 
    energy with a higher curvature, leading to increased accuracy with Laplace's method 
    (see \ref{sec:laplace-error}) and lower propagated errors from the score matching model.
    \textbf{d-e)} Free energy prediction with $\hat{\alpha}(\X)$
    gives large predicted and observed errors, reaching 8.5 meV/atom at 2500 K, 
    consistent with a) and the low curvature of $\mathcal{F}(\beta|\alpha)$.
    \textbf{f-g)} Use of $\hat{\alpha}(\X,\mom)$ brings back the low 
    predicted and observed errors required for phase prediction, 
    within the target range of 1-2 meV/atom (1.5 meV/atom at 2500 K).}\label{fig:A15}
\end{figure}

While the ensemble used for A15 predictions of Mo, figure \ref{fig:Mo-A15}, gave 
good predictions of NVT free energies using a harmonic reference potential 
in the isosurface function $\hat{\alpha}(\X)$, the performance for W was poorer, 
reaching 8.5 meV/atom at 2500 K as shown in figure \ref{fig:A15}d-e). Importantly, 
D-DOS error estimates are similarly large, meaning our estimator is
correctly indicating the isosurface function is poorly chosen in this case.\\

As discussed in \ref{sec:laplace-error}, application of Laplace's method 
requires minimizing an estimate $\mathcal{F}_{\w;\T}(\beta|\alpha)$ with respect to $\alpha$, equation (\ref{modelF}), which will be more robust to noise if the $\alpha$-curvature $\partial_\alpha^2\mathcal{F}(\beta|\alpha)$ is higher. 
We thus expect isosurface functions which give a higher curvature in $\mathcal{F}_{\w;\T}(\beta|\alpha)$ will have lower predicted and observed errors. 
In figure \ref{fig:A15}b) we show $\mathcal{F}(\beta|\alpha)$ for a 
given potential $\w_1\in\mathcal{W}_{\rm W}$ using $\hat{\alpha}(\X)$, which employs 
a harmonic potential as defined in (\ref{isosurface-H}), 
and the momentum-dependent isosurface $\hat{\alpha}(\X,\mom)$, which uses a general 
interatomic potential as defined in (\ref{isosurface-P}). As can be seen 
in \ref{fig:A15}b) for $\w=\w_1$ and in \ref{fig:A15}c) across the whole ensemble, 
the momentum-dependent isosurface gives significantly higher curvature with $\alpha$.  
Using $\hat{\alpha}(\X,\mom)$ for sampling then gives significantly lower 
predicted and observed errors, remaining within the 1-2 meV/atom limit (1.5 meV/atom at 2500 K)
required for phase stability.

\begin{figure}[!t]
    \includegraphics[width=0.99\linewidth]{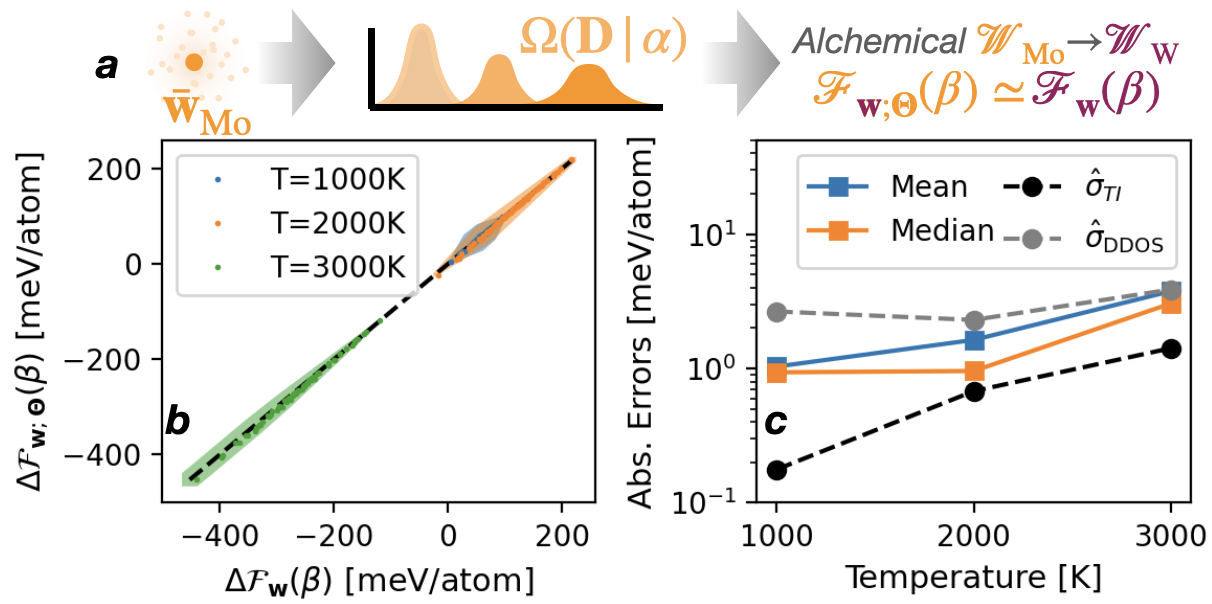}
    \caption{Alchemical transferability. \textbf{a)} D-DOS sampling with $\hat{\alpha}({\bf X})$, equation (\ref{isosurface-H}), built with reference Mo potential $\bar{\w}_{\rm Mo}$.
    However, the testing ensemble is for W, $\w\in\mathcal{W}_{\rm W}$, as shown in 
    figure \ref{fig:BCC}.
    \textbf{b)} Parity plot at three temperatures with propagated errors as in 
    other ., with median accuracy of 2 meV/atom over a 600 meV/atom range. 
    \textbf{c)} Mean and median absolute 
    errors over $\mathcal{W}_{\rm W}$. Predicted D-DOS errors propagated 
    from score matching (\ref{sec:error_analysis}) tightly bound true errors.}\label{fig:BCC-alchem}
\end{figure}

\subsection{Alchemical transferability}\label{sec:alchem}
As a final stringent test of D-DOS sampling in forward propagation, 
figure \ref{fig:BCC-alchem} shows predictions of the same 
bcc NVT free energies for the W ensemble $\mathcal{W}_{\rm W}$ 
presented in figure \ref{fig:BCC}, but now using a D-DOS 
estimator with $\hat{\alpha}(\X)$ built using the Hessian from 
a representative potential $\bar{\w}_{\rm Mo}$ from the Mo ensemble $\mathcal{W}_{\rm Mo}$. 
Despite the significant increase in explicit anharmonicity due to the 
change of reference system (\ref{fig:BCC-alchem}b), we see that 
performance is only slightly reduced, with errors remaining at 1-2 meV/atom 
(\ref{fig:BCC-alchem}c), 
within the target range of our state-of-the-art thermodynamic calculations. 
These results provide compelling evidence that the approach outlined here 
opens many perspectives for `universal' or alchemical sampling, which we discuss 
further in \ref{sec:conclusions}. 
\begin{figure}[!t]
    \includegraphics[width=0.99\linewidth]{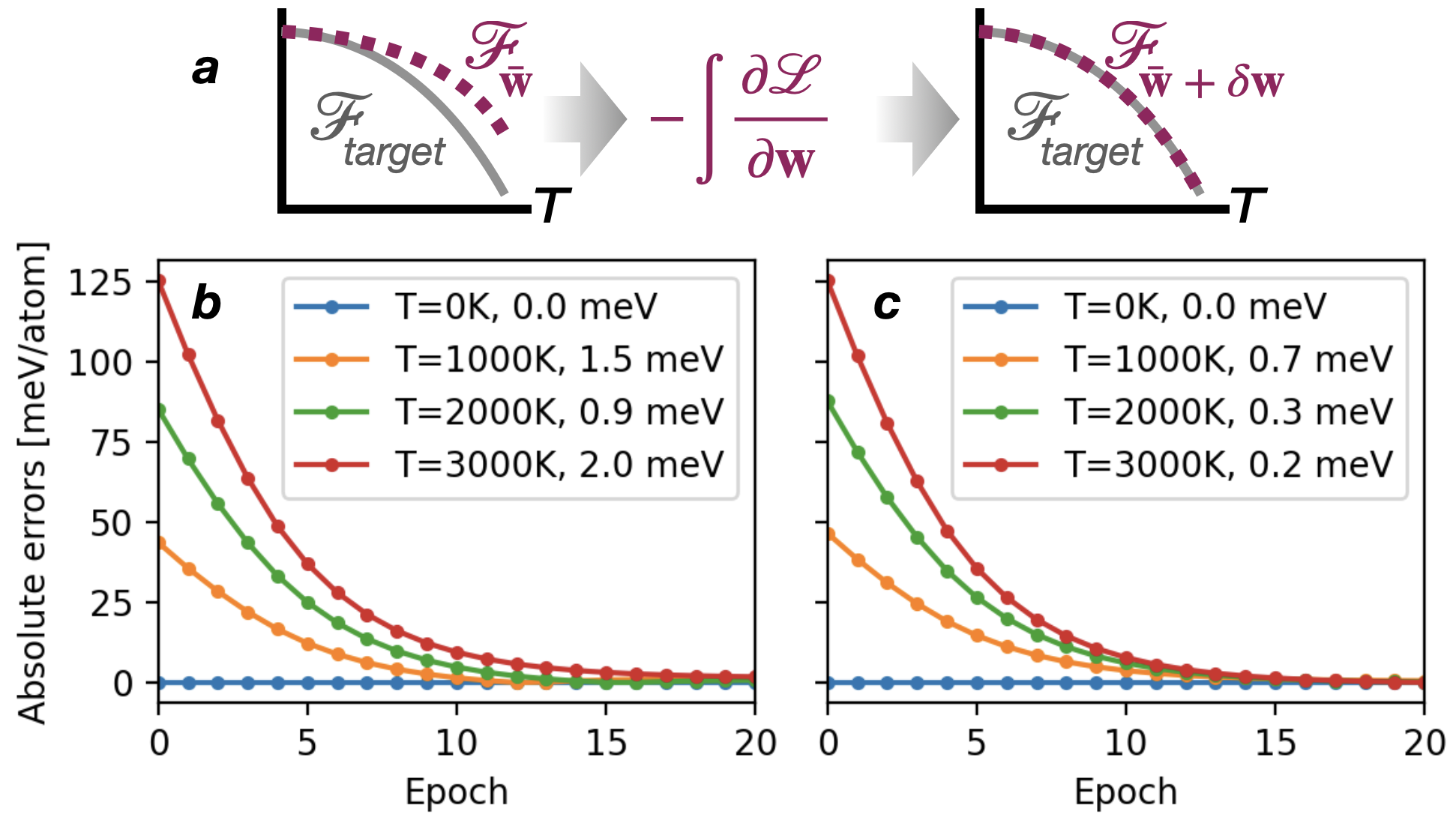}
    \caption{Fine-tuning potentials to match target W bcc NVT free energy values. 
    \textbf{a)} Access to parameter gradient of free energy allows for inclusion 
    of finite temperature properties in potential training. 
    \textbf{b-c)} Error against four target free energy values
    of single $\bar{\w}+\delta\w$ trajectory under gradient descent. 
    \textbf{b)} Target set to the 15th percentile in free energy
    over $\mathcal{W}_{\rm W}$ at each temperature. 
    \textbf{c)} Target values calculated with a single model 
    from $\mathcal{W}_{\rm W}$, a specified problem with lower errors.
    }\label{fig:NVT-inverse}
\end{figure}
\subsection{Targeting of NVT free energies}\label{sec:nvt-target}
In the next two subsections, we turn to back-propagation of parameter
variations, a long-standing goal of atomic simulations which, to the best  
of our knowledge, has never been achieved for complex thermodynamic quantities such as 
the free energy, which cannot be expressed as a simple expectation.

We first consider the `fine-tuning' of an initial interatomic potential parameter 
$\bar{\w}$, targeting some reference NVT free energies. While we
consider arbitrary targets, in applications one would target 
\textit{ab initio} values obtained from a multi-stage stratified sampling 
scheme\cite{castellano2024machinelearningassistedcanonical}, to enforce 
consistency during the training of a general purpose MLIP. 

In our inverse design procedure, we start with some regularization term $\mathcal{L}_0(\w)$, 
i.e. the training loss or the negative log likelihood from Bayesian inference\cite{swinburne2025}. Our primary target is a set of NVT free energies 
$\{\mathcal{F}(\beta_t)\}$ at a range of inverse temperatures $\{\beta_t\}$, 
giving an objective function
\begin{equation}
    \mathcal{L}(\w)
    =
    \sum_t\|\mathcal{F}_{\w;\T}(\beta_t)-\mathcal{F}(\beta_t)\|^2
    +
    r \mathcal{L}_0(\w),
    \label{target-loss}
\end{equation}
where $r$ controls the regularization strength. Access to the 
parameter gradient (\ref{gradient}) allows minimization through 
simple application of gradient descent, updating parameters through
$\w_{n+1} = \w_{n}-\delta \nabla_\w\mathcal{L}(\w_n)$.\\

Figure \ref{fig:NVT-inverse} shows the
result of this process for two different target free energies; in 
figure \ref{fig:NVT-inverse}b), 
the target is simply the 15th percentile value at each temperature across 
the W ensemble $\mathcal{W}_{\rm W}$, showing maximum error of 2 meV/atom at 3000 K.  
In this case, the target is \textit{misspecified}, i.e. it is 
not clear that a single choice of MLIP parameters is able to match the target 
value, mimicking the realistic design case where $\{\mathcal{F}(\beta_t)\}$ is 
obtained from \textit{ab initio} data. Nevertheless, our minimization approach 
smoothly converges to within 2 meV/atom at high temperature. 
Figure \ref{fig:NVT-inverse}c) shows a \textit{specified} target, using the free 
energies of a single model calculated through thermodynamic integration, which has 
similar deviation from the free energies of $\bar{\w}$ as the first target. In this 
case, we obtain essentially perfect agreement (within the sampling uncertainty) of less than 1 meV/atom at all temperatures. These results represent a second key result 
of this paper, a demonstration that finite temperature material properties 
can be included in the objective functions for negligible additional cost; 
as mentioned above, evaluation of the free energy and its gradient  
requires only microseconds of CPU effort, with no additional atomistic sampling.

\begin{figure}[!t]
    \includegraphics[width=0.99\linewidth]{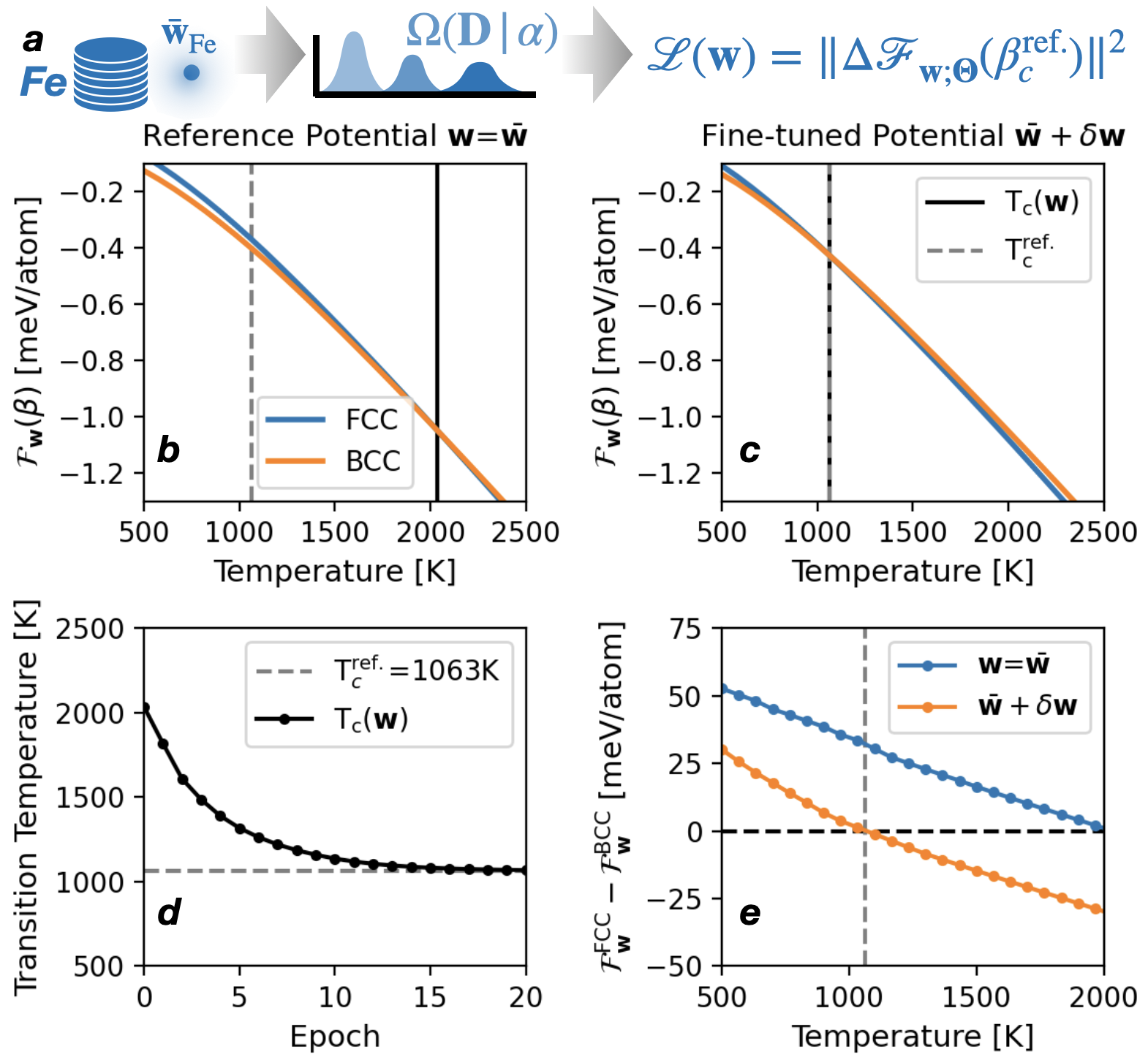}
    \caption{Inverse fine-tuning to target the $\alpha\to\gamma$ transition in Fe. 
    a) From multiphase training data in Fe, we perform D-DOS sampling with 
    $\hat{\alpha}({\bf X})$, using $\bf H$ from the reference 
    Fe potential $\bar{\w}_{\rm Fe}$.
    b) NVT free energies in fcc and bcc phases of $\bar{\w}_{\rm Fe}$, using equilibrium volume at 
    $T^{\rm ref}_c=1063$ K. The observed transition occurs at $T_c(\bar{\w})=2030$ K. 
    c) The fine-tuned potential $\bar{\w}+\delta\w$ has  the correct transition temperature. 
    d) The transition temperature smoothly decays during fine-tuning. 
    e) Free energy difference between fcc and bcc phases with temperature for 
    $\w=\bar{\w}$  (blue) and $\w=\bar{\w}+\delta\w$ (orange).
    Despite the small (40 meV/atom) change in free energy, the transition temperature is 
    reduced by nearly 1000 K, demonstrating the sensitivity of phase stability to small 
    changes in interatomic potential parameters.}\label{fig:NVT-BCC-FCC}
\end{figure}

\subsection{Targeting $\alpha\to\gamma$ transition temperature in Fe}\label{sec:alpha-gamma-tr}
As a final example, we demonstrate how back-propagation allows for 
the targeting of phase transition temperatures, to our knowledge, 
a unique ability of the D-DOS procedure. As above, targets could be calculations from established schemes\cite{castellano2024machinelearningassistedcanonical} to enforce consistency during general purpose potential training, prescribed from higher level simulations to enforce consistency within multi-scale models\cite{chen2022classical} or to experimental data, in top-down training schemes\cite{thaler2022deep}.

Our demonstration targets the bcc-fcc, or $\alpha\to\gamma$, transition in Fe. 
While known to be due to the loss of ferromagnetic ordering\cite{ma2017dynamic},
in this example of back-propagation we employ non-magnetic \texttt{SNAP} models of Fe.
Our starting point is a previously published set of \texttt{SNAP} parameters
$\bar{\w}_{\rm Fe}$\cite{goryaeva2021efficient} which was designed primarily for the simulation of point and extended defects in the bcc phase. While the model supports 
a stable fcc phase, the initial $\alpha\to\gamma$ transition temperature 
is over 2000 K, significantly above the target value of 1063 K.\\

Using $\bar{\w}_{\rm Fe}$ to generate harmonic isosurface functions 
$\hat{\alpha}(\X)$, equation (\ref{isosurface-H}), for fcc and bcc phases, 
a D-DOS score matching campaign produced estimators for 
both phases over a small range of atomic volumes, allowing calculation
of NPT free energies $\mathcal{G}_{\w;\T}(\beta)$, as detailed in 
section \ref{sec:TI}, equation (\ref{gibbsF}). With estimators 
for both the fcc and bcc free energies at the desired critical temperature 
${\rm T}_c=1063$K, the objective function reads
\begin{equation}
    \mathcal{L}(\w)
    =
    \|
    \mathcal{G}_{\w;\T}(\beta_c,{\rm bcc})
    -
    \mathcal{G}_{\w;\T}(\beta_c,{\rm fcc})
    \|^2
    +
    r \mathcal{L}_0(\w),
\end{equation}
where $\beta_c=1/({\rm k_BT}_c)$ and as in 
(\ref{target-loss}) 
$\mathcal{L}_0(\w)$ is the training loss function with a regularization 
parameter $r\geq0$ and $\beta_c=1/{\rm k_BT}_c$. Clearly, minimization of the 
first term will ensure a predicted phase transition at $\rm T_c$ as the Gibbs 
free energies are equal. As in the previous section, parameters are updated 
by gradient descent, using (\ref{gradient}) to evaluate $\nabla_\w\mathcal{L}(\w)$
by the chain rule.\\

As shown in figure \ref{fig:NVT-BCC-FCC}, this inverse fine-tuning procedure 
enables finding the subtle changes in potential parameters required 
to reproduce the desired phase boundary, 
reducing the $\alpha\to\gamma$ transition temperature from 2030 K to 1063 K,
with the gradient descent smoothly converging, figure \ref{fig:NVT-BCC-FCC}d). 
While the total free energy changes were relatively small, on the order of 
30 meV/atom, figure \ref{fig:NVT-BCC-FCC}e) demonstrates the consequence on 
the phase boundary: as the free energy gradient with temperature 
is only around 0.03 meV/atom/K, a change of 30 meV/atom results in a 1000 K
change in phase transition temperature.
\section{Discussion}\label{sec:conclusions}
We end this paper with a brief outline of the 
many perspectives for model agnostic sampling with 
D-DOS approach, within condensed matter physics
and materials science and also more widely,
many of which will be investigated in future work.

\subsection{Liquid and dynamically unstable phases}
The momentum-dependent isosurface function (\ref{isosurface-P}) 
only requires the ability to perform NVE dynamics 
with the reference potential $E_0(\X)$ and can thus be applied to 
evaluate free energies of liquids or dynamically unstable solids\cite{Park2024}.
Section \ref{sec:influence-score} demonstrated that use of (\ref{isosurface-P}) 
significantly improved prediction accuracy  when applied to the strongly anharmonic A15 phase in W. 
Of equal importance, the predicted error bounds also significantly reduced. 
A full study of liquid free energies and thus melting transitions
within the D-DOS approach will be the subject of a forthcoming study. 
For dynamically unstable (entropically stabilized) systems the 
conditional descriptor entropies may be multi-modal, potentially
requiring more advanced score models employing e.g. low-rank 
tensor approximations\cite{cui2023scalable,darby2023tensor,sherman2020estimating}
or neural-networks\cite{song2020sliced}, beyond the simple low-rank score model employed here.
We anticipate that integrating these more flexible density estimation models
will open many perspectives for sampling complex material systems. 

\subsection{Fine-tuning of foundational MLIPs}
Section \ref{sec:mlip-types} discussed the applicability of D-DOS
beyond the linear MLIPs employed in our numerical experiments (section \ref{sec:snap}). 
In particular, the D-DOS scheme is well-suited to the fine-tuning of
message passing neural network (MPNN) interatomic potentials\cite{musaelian_allegro_2023, batzner2022, batatia2022mace, cheng2024cartesian}, 
which have gained prominence due to their `foundational' ability to approximate 
broad regions of the periodic table\cite{batatia2023foundation, Bochkarev2024}. 
While MPNN training is a deep learning procedure involving millions of model weights, 
fine-tuning schemes typically only target a small number of parameters in the MPNN readout layer, 
which can be expressed in the general linear form (\ref{linearE}).
For example, uncertainty quantification schemes designed for linear models 
have recently been applied to e.g. the \texttt{MACE-MPA-0} foundation 
model\cite{perez2025uncertaintyquantificationmisspecifiedmachine}. The ability 
to fine-tune foundation models on available phase diagram data, 
either from experiment or from reference calculations, is actively sought 
in atomic modelling and will be investigated in the near future. In particular,
the light computational demand and storage requirements of D-DOS mean 
production of a `foundational' D-DOS sampling library for foundational MPNNs 
is feasible and indeed desirable, allowing practitioners to rapidly assess how 
e.g. fine-tuning on additional training data improves free energy predictions.\\

Beyond physics models, a broad range of computational tasks reduce 
to evaluating high-dimensional integrals. We here highlight two well known 
examples which can be cast to the generalized linear form (\ref{linearE}) 
amenable to D-DOS estimation.

\subsection{Generalized Ising Models}
Generalized Ising models are widely used in condensed matter physics and materials 
science. For example, the popular cluster expansion\cite{Erhart2024}
model is used to approximate the configurational entropy of multi-component lattice systems. 
Cluster expansion models have energies of the form (\ref{linearE}), where descriptor features 
$\tD$ (typically denoted as $\boldsymbol{\Gamma}$ in the cluster expansion literature) are correlation functions of some multi-component lattice and $\w$ (typically denoted as $\bf J$)
is fit to \textit{ab initio} data. The central difference is that the configuration
space is no longer a vector $\X\in\mathbb{R}^{3N}$ but $\boldsymbol{\sigma}\in[1,S]^N$, 
the discrete set of all $S$-component configurations across $N$ lattice sites. Generalized Ising models have a rich phenomenology,
in particular second order phase transitions, and the development of efficient sampling methods is an active area of research\cite{zhang2020loop,Marchand2023}. 
However, we anticipate 
that the D-DOS approach could be applied in some settings, 
e.g. for model-agnostic sampling of equilibrium averages central to 
the cluster expansion models\cite{Erhart2024}, with the same advantages for uncertainty quantification and inverse design as shown for atomistic systems.\\

\subsection{Probabilistic learning and Bayesian inference}
Probabilistic machine learning and Bayesian inference both require 
fast evaluation of high dimensional integral\cite{von2011bayesian,bhat2010derivation,lotfi2022bayesian}.
A quite direct analogy with D-DOS can be found if we equate $\X$ with 
model parameters, $\w$ with hyperparameters and the free energy with e.g. 
the log evidence per parameter or the log of the posterior predictive distribution,
in the limit of a very large number of model parameters\cite{duffield2024scalable}.
In this setting the common Laplace approximation used to evaluate integrals 
in Bayesian inference exactly corresponds here to the harmonic approximation
used for the construction of $\hat{\alpha}(\X)$, equation  (\ref{isosurface-H}). 
We can thus apply the D-DOS approach if we can write the log likelihood
as the sum of the harmonic expansion used in the Laplace approximation and a 
generalized linear feature model $\w\cdot\tD$. In this setting the model-agnostic estimation 
of D-DOS would allow rapid, differentiable exploration of the hyperparameter 
dependence on model predictions actively sought in Bayesian machine learning\cite{blondel2022}.\\

\section{Conclusions}
In this study, we have revised traditional methods to estimate the vibrational free energy 
of atomic systems by proposing a novel approach that leverages 
descriptors, high-dimensional feature vectors of atomic configurations. 
We introduced the foundations for estimating 
key quantities of interest in 
this high-dimensional descriptor space, such as the descriptor density of states and corresponding entropy.
{The present reformulations introduce a model-agnostic sampling scheme for atomic simulation}. \\

Rather than existing methods which return
free energy estimates for a specific value of interatomic potential parameters, we instead return an estimator that can predict free energies 
over a broad range of model parameters. This is a significant change 
in approach that not only allows for rapid forward propagation of 
parameter uncertainties to finite temperature properties, but also
uniquely allows for inverse fine-tuning of e.g. phase boundaries 
through back-propagation, both long-desired capabilities 
in computational materials science.

Central to our scheme is the \textit{descriptor density of states} (D-DOS),
a multidimensional generalization of the energy density of states. 
We showed how score matching the descriptor entropy (log D-DOS) enables 
free energy estimation without numerical integration for the wide class 
of interatomic potentials that can be expressed as a linear model 
of descriptor features. A large range of tasks in computational 
science reduce to evaluating high-dimensional integrals, and thus 
many can be cast in a form directly analogous to the free energy estimation problem D-DOS solves, some of which we 
outlined above. 
We fully anticipate that more examples can be found which would allow the application of the D-DOS approach to a wide range of computational tasks. 

\section{\label{sec:acknowledgements} Acknowledgments}
We gratefully acknowledge the hospitality of the Institute for Pure and 
Applied Mathematics at University of California, Los Angeles, 
the Institute for Mathematical and Statistical Innovation at the 
University of Chicago and the Institut Pascal at Université Paris-Saclay,  
which is supported by ANR-11-IDEX-0003-01. TDS gratefully acknowledges 
support from ANR grants ANR-19-CE46-0006-1, ANR-23-CE46-0006-1, IDRIS allocation 
A0120913455 and an Emergence@INP grant from the CNRS. All authors acknowledge the support 
from GENCI - (CINES/CCRT) computer centre under Grant No. A0170906973.

\section{Data Availability}
After peer review, an open source code repository of our score matching scheme, 
enabling D-DOS estimation for any descriptor implemented in 
the \texttt{LAMMPS}\cite{LAMMPS} molecular dynamics code, will be available
at \url{www.github.com/tomswinburne/DescriptorDOS.git}.

\newpage

\onecolumngrid
\appendix
\section{Summary of Laplace's method}
\label{app:laplace}
Laplace's method, also known as the steepest descents method, is a well-known identity allowing the evaluation of an integral of an exponentiated function multiplied by a large number. We provide a brief summary of the method here, and we refer the reader to e.g. \citep{wong2001asymptotic} for further information.
The method applies to a function $f({\bf x})$, ${\bf x}\in\mathbb{R}^n$ which is twice differentiable.
We partition the domain $\mathbb{R}^n=\cup_{l=1}^L\mathcal{R}_l$ into regions $\mathcal{R}_l$ each with a single maximum ${\bf x}^*_l$, where 
the negative Hessian matrix ${\bf H}_l=-\nabla_{\bf x}\nabla^\top_{\bf x}f_{|\mathcal{R}_l} \in\mathbb{R}^{n\times n}$ of $Nf({\bf x})$ has $\mathcal{O}(n)$ positive eigenvalues $\lambda_p\geq0$, no negative eigenvalues, and all entries of ${\bf H}_l$ (and thus all $\lambda_p$) are independent of $N$. 
In the limit $N\to\infty$ the integral in $\mathcal{R}_l$ is dominated 
by the maximum ${\bf x}^*_l$. Proof of Laplace's method uses Taylor expansions 
of $f({\bf x})$ around ${\bf x}^*_l$ to provide upper and lower bounds, which in the limit $N\to\infty$ both converge to the same Gaussian integral, giving 
\begin{equation}
    \lim_{N\to\infty}
    \int_{\mathbb{R}^n}
    \exp[Nf({\bf x})]{\rm d}{\bf x}
    =
    \sum_{l=1}^L
    \frac{\exp[Nf({\bf x}^*_l)]}
    {\sqrt{(2\pi N)^n|{\bf H}_l|}}.
\end{equation}
In the case of constant $n$ as $N\to\infty$ it is simple to show that 
the limiting form of the log integral is 
\begin{equation}
    \lim_{N\to\infty}
    \frac{1}{N}\ln
    \int_{\mathbb{R}^n}
    \exp[Nf({\bf x})]{\rm d}{\bf x}
    =
    \max_l 
    f({\bf x}^*_l).
\end{equation}
where we use the fact that 
\begin{equation}
    \lim_{N\to\infty}
    \frac{1}{N}\ln
    \sqrt{(2\pi N)^n|{\bf H}_l|}
    =
    \lim_{N\to\infty}
    \frac{1}{2N}\left(n\ln|2\pi N| + \sum_{p=1}^n \ln\lambda_p\right)
    =0.
\end{equation}
as $\lim_{N\to\infty}n/N=0$ and $\lim_{N\to\infty}(1/N)\ln|N|=0$.
When the argument of the function has dimension which scales with 
$N$, i.e. $n=rN$, $r>0$, ($r=3$ for Hessians), the above simplification does 
not hold. In general, the integral will depend on higher order gradients 
to correctly take the limit. Note that the above is
distinct from the common use of Laplace's method to approximate the partition 
function integral $\int_{\mathbb{R}^{3N}} \exp[-\beta E({\bf X})]{\rm d}{\bf X}$; although the 
dimension of $\bf X$ is extensive, in this case Laplace's method is used in the 
low temperature limit $\beta\to\infty$, rather than $N\to\infty$.
\section{Isosurface for a harmonic solid}\label{app:isosurface_harmonic}
\subsection{Sampling}
For solid systems we use a harmonic reference potential energy $E_0(\X)$
and isosurface function $\hat{\alpha}({\bf X})$
\begin{equation}
    E_0(\X)\equiv\frac{[\X-\X_0]^\top{\bf H}[\X-\X_0]}{2N},\quad 
    \hat{\alpha}(\X)
    \equiv 
    \ln\left|\frac{\hat{E}_0({\bf X})}{{\rm U}_0}\right|.
\end{equation}
We assume that ${\bf H}$ has $3N'=3N-3$ positive eigenvalues $\nu_l>0$, $l>3$
with normalized eigenvectors ${\bf v}_l$. We can thus define normal mode coordinates
$\tilde{\rm X}_l\equiv {\bf v}_l\cdot\X/\sqrt{\nu_l}$, $l>3$. 
In addition, we have $3$ zero modes $\nu_{l}=0$, $l=1,2,3$ with eigenvectors 
selecting the center of mass $\bar{\bf x}$ multiplied by $\sqrt{N}$, i.e. 
$\tilde{\rm X}_l\equiv \sqrt{N}\bar{\rm x}_l$, $l=1,2,3$, which 
meaning we can always ensure normal modes have zero net displacement, 
i.e. enforce $ {\bf v}_l\cdot{\bf 1}=0$, $l>3$.\\

In normal mode coordinates, the energy writes
\begin{equation}
    E_0(\X)= \frac{1}{2N}\sum_{l=4}^{l=3N} {\nu_l} \|{\bf v}_l\cdot\X\|^2 = \frac{1}{2N}\sum_{l=4}^{l=3N} \tilde{\rm X}^2_l = \frac{\tilde{\rm R}^2}{2N}.
\end{equation}
Sampling the isosurface $\hat{\alpha}(\X)=\alpha$ is clearly equivalent to sampling $E_0(\X)={\rm U}_0\exp(\alpha)$, which in normal mode coordinates amounts to sampling the surface
of a hypersphere with radius $\tilde{\rm R}=\sqrt{2N{\rm U}_0}\exp(\alpha/2)$.\\

With a unit vector ${\bf u}=[{\rm u}_1,\dots,{\rm u}_{3N'}]\in\mathbb{R}^{3N'}$ on the $3N'$ dimensional hypersphere, isosurface samples can then be produced through 
\begin{equation}
    \X_\alpha[{\bf u}] \equiv \X_0
    +
    \sum_{l=1}^{3N'}
    \sqrt{\frac{2N{\rm U}_0\exp(\alpha)}{\nu_l}}{\rm u}_l{\bf v}_l
    ,\quad\Rightarrow\quad
    E_0\left(\X_\alpha[{\bf u}]\right)
    =
    {\rm U}_0\exp(\alpha)
    \sum_{l=1}^{3N'}
    {\rm u}^2_l
    =
    {\rm U}_0\exp(\alpha).
\end{equation}
Importantly, the sampling procedure can be trivially parallelized as 
we can generate independent samples $\{{\bf u}\}$ on each parallel worker,
providing each worker with a unique seed for pseudo-random number generation.
\subsection{Isosurface volume and isosurface entropy}
For harmonic isosurface functions $\hat{\alpha}(\X)$, we can express the isosurface volume (\ref{volume}) in normal mode coordinates using standard expressions for change of variables:
%is defined as 
\begin{align}
    \Omega(\alpha) &= \int_{\mathbb{R}^{3N}}
    \delta(\hat{\alpha}(\X)-\alpha)
    {\rm d}{\X}\\
    &=
    \frac{V}{\prod_{l=4}^{3N}\sqrt{\nu_l}}
    \int_{\mathbb{R}^{3N'}}
    \delta\left(
    \ln\left|\sum_{l=4}^{3N}
    \tilde{\rm X}_l^2/(2N)\right|-\alpha
    \right)
    \prod_{l=4}^{3N}{\rm d}\tilde{\rm X}_l
    .
\end{align}
Converting to spherical coordinates we find, using the expression for the surface area of 
a unit sphere in $3N'$ dimensions as $S_{3N'}=2{\pi}^{3N'/2}/\Gamma(3N'/2)$, 
then again changing variables with ${\rm d}\tilde{\rm R}=\sqrt{N{\rm U}_0/2}
\exp(\alpha/2){\rm d}\alpha$, we find that
\begin{align}
    \Omega(\alpha)
    &=
    \frac{2V{\pi}^{3N'/2}}{\Gamma(3N'/2)}
    \int_{\mathbb{R}_+}
    \delta\left(
    \ln\left|\tilde{\rm R}^2/(2N)\right|-\alpha
    \right)
    \tilde{\rm R}^{3N'-1}
    {\rm d}\tilde{\rm R}\\
    &=
    \frac{V\sqrt{\rm U}^{3N'}_0}{\prod_{l=4}^{3N}\sqrt{\nu_l}}
    \frac{\sqrt{2\pi N}^{3N'}}{\Gamma(3N'/2)}
    \exp(3N\alpha/2).
\end{align}

We thus see that the isosurface volume for harmonic solids has the general form
\begin{equation}
\Omega(\alpha) = \Omega_0\exp(3N\alpha/2),\quad \Omega_0=\frac{V\sqrt{\rm U}^{3N'}_0}{\prod_{l=4}^{3N}\sqrt{\nu_l}}\frac{\sqrt{2\pi N}^{3N'}}{\Gamma(3N'/2)},
\end{equation}
giving an isosurface entropy
\begin{equation}
    \mathcal{S}_0(\alpha)
\equiv
\lim_{N\to\infty}\frac{1}{N}\ln|\Omega(\alpha)/\lambda^{3N}_0(\beta)|
=
S_0+3\alpha/2,
\quad
S_0 = \ln|\lambda^{3}_0(\beta)|+(1/N)\ln|\Omega_0|.
\label{app_iso_entropy}
\end{equation}
While we can simplify the expression for the $\ln\Omega_0$ using Stirling's approximation 
we shall see this is not required. 
\subsection{Isosurface entropy and connection to harmonic free energy}
Using standard Gaussian integrals, the partition function of a harmonic system reads,
with $\lambda_0(\beta)=h\sqrt{\beta/(2\pi m)}$,
\begin{equation}
Z_0(\beta)=
    \frac{1}{\lambda^{3N}_0(\beta)}
    \int_{\mathbb{R}^{3N}}
    \exp[-N\beta E_0(\X)]{\rm d}\X =
    \frac{V}{\lambda^{3N}_0(\beta)}
    \prod_{l=4}^{3N}
    \frac{1}{\sqrt{2\pi\beta\nu_l}},
\end{equation}
giving a free energy in the limit $N\to\infty$
\begin{equation}
    \mathcal{F}_0(\beta)
    \equiv 
    \lim_{N\to\infty}
    \frac{-1}{N\beta}\ln|Z_0(\beta)|
    =
    \frac{1}{N\beta}\sum_{l=4}^{3N}\ln\left|\beta\hbar\sqrt{\nu_l/m}\right|.
\end{equation}
We can also write $\mathcal{F}_0(\beta)$ using the isosurface entropy
defined in equation (\ref{app_iso_entropy}) and applying Laplace's method, i.e. 
\begin{align}
    \beta\mathcal{F}_0(\beta)
    &=
    \lim_{N\to\infty}
    \frac{-1}{N}
    \ln\left|
    \frac{1}{\lambda^{3N}_0(\beta)}
    \int_{\mathbb{R}}
    \Omega(\alpha)
    \exp(-N\beta{\rm U}_0e^\alpha){\rm d}\alpha\right|\\
    &=
    \lim_{N\to\infty}
    \frac{-1}{N}
    \ln\left|
    e^{NS_0}
    \int_{\mathbb{R}}
    \exp(3N\alpha/2-N\beta{\rm U}_0e^\alpha){\rm d}\alpha\right|,\\
    &=\min_\alpha {\rm U}_0e^\alpha - 3\alpha/2-S_0,\\
    &= {\rm U}_0e^\alpha - 3\alpha/2-S_0\Big|_{\alpha = -\ln|2\beta{\rm U}_0/3|},\\
    &= 3/2 - S_0 + 3/2\ln|2\beta{\rm U}_0/3|,\\
    \Rightarrow
    S_0&=3/2 + 3/2 \ln|2\beta{\rm U}_0/3|-\beta\mathcal{F}_0(\beta)
\end{align}
which allows us to express the constant $S_0$ purely in terms of the harmonic free energy. 

\section{Momentum-dependent isosurface}\label{app:isosurface_momentum}
Estimating the free energy of e.g. liquid or highly anharmonic phases typically 
requires more complex reference potential energy models than the harmonic 
form (\ref{isosurface-H}). While we leave a comprehensive numerical study for future work, the following details how the treatment in \ref{sec:CDDOS} 
can be generalized to a momentum dependent isosurface
\begin{equation}
    \hat{\alpha}(\X,\mom)
    \equiv 
    \ln\left|\frac{\hat{K}({\bf P}) + \hat{E}_0({\bf X})}{{\rm U}_0}\right|.
\end{equation}
using an intensive kinetic energy function $\hat{K}(\mom)=(1/N)\sum_{i=1}^{3N}{\rm p}^2_i/(2m_i)$. Isosurface sampling then corresponds to microcanonical 
(NVE) dynamics with any reference potential, where the per-atom internal energy satisfies 
$\mathcal{U}={\rm U}_0\exp(\alpha)$. Such a generalization has close analogies 
with Hamiltonian Monte Carlo methods\cite{betancourt2017conceptual},
which can use generalized kinetic energies\cite{livingstone2019kinetic}.
The isosurface volume of $\hat{\alpha}(\X,\mom)=\alpha$ is defined as
\begin{equation}
    \Omega(\alpha)
    \equiv
    \int_{\mathbb{R}^{3N} \times \mathbb{R}^{3N}}
    \delta\left(\hat{\alpha}(\X,\mom)-\alpha\right)
    {\rm d}{\X}{\rm d}{\mom}.
    \label{Omega_alpha_k}
\end{equation}
and we evaluate the entropy below. In this case, we treat $K$ as an additional descriptor 
to give an extended conditional descriptor density of states
\begin{align}
    \Omega(\tD\oplus K|\alpha)
    \equiv
    &
    N
    \int_{\mathbb{R}^{3N} \times \mathbb{R}^{3N}}
    \frac{\delta(\hat{K}(\mom)-K)
    \delta\left(\hat{\alpha}(\X,\mom)-\alpha\right)}
    {\Omega(\alpha)}
    \nonumber\\
    &\times
    \delta\left(
    N\tD-\sum_{i=1}^N \hat{\bphi}(\D_i({\bf X}))
    \right)
    {\rm d}{\X}
    {\rm d}{\mom},
\end{align} 
By the same manipulations as for the momentum-independent case, this extended 
conditional descriptor density of states is normalized:
\begin{equation}
    \int_{\mathbb{R}^D \times \mathbb{R}_+}\Omega(\tD\oplus K|\alpha){\rm d}\tD{\rm d}K=1.
\end{equation}

However, as samples 
are not independent, the efficacy will depend on the decorrelation time\cite{rousset2010free} 
of microcanonical trajectories. A full study of how such momentum-dependent isosurfaces can be 
used to estimate the descriptor density of states $\Omega(\tD)$
and thus the free energy of liquid phases and melting temperatures
will be the focus of future work.
\subsection{Isosurface entropy}
The isosurface entropy $\mathcal{S}_0(\alpha)$ cannot be evaluated 
analytically and instead requires free energy estimation schemes such as thermodynamic integration, discussed in \ref{sec:FE}. To see how this emerges, 
we use the definition of the isosurface entropy (\ref{entropy}) to write the free energy as 
\begin{align}
    \beta\mathcal{F}_0(\beta) &=
    \lim_{N\to\infty}
    \frac{-1}{N}\ln\left|\frac{1}{h^3N}
    \int_{\mathbb{R}} \Omega(\alpha)\exp(-N\beta\exp(\alpha)) d \alpha
    \right|,\\
    &=
    \lim_{N\to\infty}
    \frac{-1}{N}\ln\left|
    \int_{\mathbb{R}} \exp(N\mathcal{S}_0(\alpha)-N\beta\exp(\alpha)) d \alpha
    \right|,\\
    &=
    \min_\alpha \beta{\rm U}_0\exp(\alpha)-\mathcal{S}_0(\alpha).
\end{align}
It is clear that this minimum is satisfied when 
$\partial_\alpha\mathcal{S}_0(\alpha)=\beta{\rm U}_0\exp(\alpha)$, and at the minimum 
${\rm U}_0\exp(\alpha)$ is clearly the internal energy $\mathcal{U}_0(\beta)$. 
We can therefore define $\beta_\alpha$ through the condition $\mathcal{U}_0(\beta_\alpha)\equiv{\rm U}_0\exp(\alpha)$ and thus write
\begin{align}
    \beta_\alpha\mathcal{F}_0(\beta_\alpha)
    &=
    \beta_\alpha\mathcal{U}_0(\beta_\alpha) - \mathcal{S}_0(\alpha).
    %\\
    %\mathcal{S}_0(\alpha)&=\beta_\alpha[\mathcal{U}_0(\beta_\alpha)-\mathcal{F}_0(\beta_\alpha)],
\end{align}

With a tabulation of the intensive per-atom free energy 
$\mathcal{F}_0(\beta)$ and total internal energy $\mathcal{U}_0(\beta)$ 
over a range of temperatures $1/\beta$, the isosurface entropy reads
\begin{equation}
    \mathcal{S}_0(\alpha) \equiv 
    \beta[\mathcal{U}_0(\beta_\alpha)-\mathcal{F}_0(\beta_\alpha)],
    \quad
    \mathcal{U}_0(\beta_\alpha)\equiv {\rm U}_0\exp(\alpha).
\end{equation}
The value of $\beta_\alpha$ is uniquely defined when $\mathcal{U}_0(\beta)$
is monotonic with $\beta$.

\section{Intensivity of the descriptor entropy}\label{app:intensive}
This appendix provides a proof that the descriptor entropy $\mathcal{S}(\tD|\alpha)$, equation (\ref{descriptor-entropy}), is intensive.\\

By the locality of the descriptor energy (\ref{descriptorE}), 
any two per-atom feature vectors $\tD_i,\tD_j$ will be 
independent when the corresponding atoms are spatially separated, 
i.e. $|{\bf r}_{ij}|\to\infty$. 
As a result, the per-atom feature vector $\tD_i$ will have nonzero correlation with only a finite number $N_c\ll N$ of other per-atom feature vectors,
indexed by some set $\mathcal{N}_i\subset\{1,\dots,N\}$,
which has strong implications for the global vector $N\tD=\sum_{i=1}^N\tD_i$. 
In particular, it is clear that any cumulant\cite{Cover2006} of $N\tD$ 
will be extensive, scaling linearly with $N$ as $N\to\infty$. 
The first cumulant is the mean $N\langle\tD\rangle_\alpha$,
where $\langle\tD\rangle_\alpha$ is clearly intensive. 
Defining $\delta\tD_i\equiv\tD_i-\langle\tD\rangle_\alpha$ and thus $\delta\tD$, the covariance of $N\tD$ writes 
%\cl{I'm quite confused with the cumulant notation below maybe you should used $\otimes$ definition as for thrid cumulant}
\begin{equation}
    N^2\langle\delta\tD\otimes\delta\tD\rangle_\alpha 
    =
    \sum_{i=1}^N\sum_{l\in\mathcal{N}_i}\langle\delta\tD_i\otimes\delta\tD_{l}\rangle_\alpha
    =
    N\bSig_\alpha
    \in\mathbb{R}^{D\times D},
\end{equation}
where $\bSig_\alpha$ is an average over each atom $i$ of the sum of $N_c$ covariance matrices between $i$ and neighbors $l\in\mathcal{N}_i$, which is manifestly intensive. The third order cumulant writes
\begin{equation}
    N^3\langle\delta\tD\otimes\delta\tD\otimes\delta\tD\rangle_\alpha
    =
    \sum_{i=1}^N
    \sum_{l\in\mathcal{N}_i}
    \sum_{m\in\mathcal{N}_l}
    \sum_{m=0}^{N_c}
    \langle\delta\tD_i\otimes\delta\tD_{l}\otimes\delta\tD_{m}\rangle_\alpha
    =
    N \boldsymbol{\Xi}_\alpha\in\mathbb{R}^{D\times D\times D}.
\end{equation}

where $\boldsymbol{\Xi}_\alpha$ is an average over each atom $i$ of the $N_c(N_c+1)/2$ third-order correlations between $i$ and neighbors $l\in\mathcal{N}_i$ and $m\in\mathcal{N}_l$. 
As before, this is manifestly intensive, and  we can continue this procedure to arbitrarily high orders.  
We can therefore define an intensive cumulant generating function \cite{Cover2006} of $\Omega(\tD|\alpha)$ in
the form 
\begin{align}
    J({\bf v}|\alpha)
    &\equiv
    \frac{1}{N}\ln\left|
    \int_{\mathbb{R}^D}
    e^{N{\bf v}\cdot{\tD}}
    \Omega(\tD|\alpha)
    {\rm d}\tD\right|\\
    &=\bmu_\alpha\cdot{\bf v}
    +
    \frac{1}{2}
    {\bf v}^\top\bSig_\alpha{\bf v}
    +\frac{1}{6}
    {\bf v}^\top\boldsymbol{\Xi}_\alpha : {\bf v}\otimes{\bf v}
    +\dots,
\end{align}
where ${\bf v}\in\mathbb{R}^D$ and $J({\bf v}|\alpha)$ are clearly intensive, meaning that we have the identity
%The interest of this cumulant generating function is to get the conditional density of states, and the descriptor entropy, using the inverse Laplace transform from this equality: 
\begin{equation}
    \int_{\mathbb{R}^D} \exp{ (N {\bf v} \cdot  \tD} )  \Omega\left(\tD | \alpha\right) d \tD = \exp{\left(N J( {\bf v} | \alpha) \right) } \, . 
    \label{laplace-1}
\end{equation}
As the cumulants of $\Omega(\tD|\alpha)$ are finite, 
we know that $\Omega(\tD|\alpha)$ has a 
global maximum at finite $\tD$. As a result, we define the descriptor entropy as the (negative) Legendre-Fenchel transform of the cumulant generating function:
\begin{equation}
    \mathcal{S}(\tD|\alpha) 
    \equiv
    \min_{\bf v} J( {\bf v} | \alpha) - {\bf v} \cdot  \tD
    \equiv 
    \lim_{N\to\infty}\frac{1}{N}
    \ln|\Omega(\tD|\alpha)|,
\end{equation}
which is clearly both intensive and convex in $\tD$.
\begin{comment}
It is possible to show the intensive nature of 
$\mathcal{S}({\tD}|\alpha)$ through application of the inverse Laplace transform. As the cumulants of $\Omega(\tD|\alpha)$ are finite, 
we know that $\Omega(\tD|\alpha)$ has a 
global maximum at finite $\tD$. We can 
then apply Laplace's method to (\ref{laplace-1}) to find that, using the definition (\ref{descriptor-entropy}) of $\mathcal{S}({\tD}|\alpha)$,
\begin{equation}
    \lim_{N\to\infty} J( {\bf v} | \alpha)=\max_{\tD} \mathcal{S}(\tD|\alpha)+\mathbf{v}\cdot\tD,
    \quad \Rightarrow\quad 
    \mathcal{S}(\tD^*_{\bf v}|\alpha) = J( {\bf v} | \alpha) - \mathbf{v}\cdot\tD^*_{\bf v},
    \label{laplace-2}
\end{equation}
where $\tD^*_{\bf v}$ solves the maximum condition. 

As $\tD$ has a dimension $D$ that is  independent of $N$,
 all prefactor terms vanish from the log integral as $N\to\infty$ (appendix \ref{app:laplace}).
\end{comment}
We have thus established that $\mathcal{S}({\tD}|\alpha)$ is intensive, as it is the sum of 
two manifestly intensive terms. As discussed in section \ref{sec:SM-FEP}, 
the conditional free energy (\ref{condF_final}) can be expressed in terms of the cumulant generating 
function, with ${\bf v}=-\beta\w$; however, as we detail, we instead use score matching to estimate higher order moments.
\section{Derivation of the score matching loss\label{app:score_matching}}
Our starting point is the definition of the score matching loss, 
the Fisher divergence\cite{hyvarinen2005estimation}
\begin{align}
    L(\T|\alpha)
    &\equiv
    \frac{N}{2}\langle\|
    \ScoreD
    -
    \nabla\mathcal{S}(\tD|\alpha)\|^2
    \rangle_\alpha
    -L_0\\
    &=
    \frac{N}{2}\langle\|
    \ScoreD
    \|^2\rangle_\alpha
    -
    \langle\nabla\mathcal{S}(\tD|\alpha)\cdot\ScoreD\rangle_\alpha,
\end{align}
where $L_0$ is a $\T$-independent constant and we set $L(\T|\alpha)$ to be $\mathcal{O}(N)$ by convention.
As isosurface averages $\hat{\alpha}(\X)=\alpha$ 
can clearly be written an integral over the D-DOS 
$\Omega(\tD|\alpha)= \exp[N\mathcal{S}(\tD|\alpha)]$, we use integration by parts to write
\begin{align}
    L(\T|\alpha)
    &\equiv
    \frac{N}{2}\langle\|\ScoreD\|^2\rangle_\alpha
    -
    N\int \nabla\mathcal{S}(\tD|\alpha)\cdot\ScoreD
    \exp[N\mathcal{S}(\tD|\alpha)]{\rm d}\tD,\\
    &\equiv
    \frac{N}{2}\langle\|\ScoreD\|^2\rangle_\alpha
    +
    \langle\nabla\cdot\ScoreD\rangle_\alpha,
\end{align}
as given in the main text.
\subsection{Analysis in the limit $N\to\infty$}
As we discuss in section \ref{sec:SM}, whilst the descriptor entropy $\mathcal{S}(\tD|\alpha)$ is intensive, 
averages of $\mathcal{S}(\tD|\alpha)$ or gradients over $\alpha$ will in general give rise to terms
inversely proportional to $N$. To correctly infer the solution in the limit $N\to\infty$, 
we make the multiscale hierarchy\cite{pavliotis2008multiscale}
\begin{equation}
    L(\T|\alpha)
    =
    \sum_s N^{1-s}
    L_s(\T|\alpha)
\end{equation}
To solve this hierarchy we can in principle define a multiscale solution $\T^{(S)}$ such that 
\begin{equation}
\nabla L(\T^{(S)}|\alpha) = {\bf 0} + \mathcal{O}(N^{-S}). 
\end{equation}
We can find a solution $\T^{(S)}$ in a recursive fashion, first minimizing 
$L_0(\T|\alpha)$ to give $\T^{(0)}$, then minimizing $L_1(\T|\alpha)$ under the 
constraint $\nabla L_0(\T|\alpha)=0$ to give $\T^{(1)}$, and so on. However, 
we only consider models where $ L(\T|\alpha)$ is linear in $\T$, meaning the 
loss gradient can be decomposed as
\begin{equation}
    \sum_s N^{1-s}\left[{\bf A}_s\T - {\bf b}_s\right] = {\bf 0}.
\end{equation}
Respecting the multi-scale hierarchy then equates to ensuring $\T^{(S+1)}-\T^{(S)}$ is in 
the null space of all ${\bf A}_s$, $s\leq S$. Whilst we investigated solving 
each term in this hierarchy independently, in practice this had negligable improvement over 
simply minmizing the score matching loss (\ref{smloss}) as the linear solve will naturally 
ensure the solution respects the multi-scale hierarchy to a within numerical tolerance. 

\section{Databases design for Mo bcc and A15 \label{db:mo}}

We have performed an iterative construction of the database. The final aim is to have a potential that satisfies the following requirements: (i) it should reproduce the \textit{ab initio} elastic constants at 0~K;
(ii) it must provide a reasonable thermal expansion from 0 ~K to the melting temperature; and (iii) it should mimic the thermodynamics of BCC and A15 phase from  
0~K to the melting temperature.

The DFT calculations were performed using \textsf{VASP}~\cite{VASP}. We have used a PAW pseudopotential~\cite{PAW}: we have used PPs with $sp$ core states and 12 valence electrons in the $4s^24p^64d^55s^1$ states. The cut-off energy for plane-waves is 500 eV. In order to sample reciprocal space, we used Monkhorst-Pack~\cite{Monkhorst1976} method to build a constant $k$-points density $\rho_k = 1 / (24 a_0)^3$ for all the computed configurations, which translates in $6 \times 6 \times 6 $ $k$-points for the 128-atom cell of bcc Mo. Methfessel and Paxton~\cite{Methfessel1989} smearing algorithm with $\sigma = 0.3$ eV is used. We have used  GGA exchange correlation in PBE~\cite{perdew1996generalized}.

Firstly, we generated a minimal \textit{ab initio} database, DB$_1$, designed to build the initial version of the potentials. These potentials were then used to generate additional configurations similar to the defects we intend to simulate. The configurations were then computed using DFT without structural relaxation and reintegrated into the more complete database, DB$_2$. 
We reiterate the procedure from DB$_2$ to DB$_3$.
All generated configurations are collected in Table~\ref{tab:dataset_all}. In the following, we detail each component of the database. In the end, the different databases are ruled by the following inclusion relations: DB$_1$ $\subset$ DB$_2$ $\subset$ DB$_3$.

\begin{table}[ht]
    \renewcommand{\arraystretch}{1.5}
  \resizebox{0.5\textwidth}{!}{
  \begin{tabular}{lcccccc|c}
  {DB$_1$} & \multicolumn{6}{c}{Temperature in K} &  {Total} \\  %\cline{2-6}
      & $0$  & $875$  & $1750$  & $2625$ & $3500$  & MDr \\  \hline
    \texttt{Cxx} & 13 & 0 & 0 & 0 & 0 & 0 & 13 \\
    \texttt{$\epsilon$\_bulk} & 1000 & 0 & 0 & 0 & 0 & 0 & 1000 \\
    \texttt{noised\_bulk} & 1 & 10 & 10 & 10 & 10 & 0 & 41 \\
    \texttt{noised\_V}$_1$ & 10 & 10 & 10 & 10 & 10 & 0 & 50 \\
\texttt{noised\_V}$_2$ & 30 & 30 & 30 & 30 & 30 & 0 & 150 \\
    \texttt{NEB\_}$V_{1,2,3}$ & 21 & 0 & 0 & 0 & 0 & 0 & 21 \\
    \hline
    Total DB$_1$ & 1075 & 50 & 50 & 50 & 50 & 0 & 1275 \\
    \hline 
    \hline 
    {DB$_2$} & \multicolumn{6}{c}{} &  {} \\
    & & & & & & \\
    \hline 
    DB$_1$ & 1075 & 50 & 50 & 50 & 50 & 0 & 1275 \\ 
    \texttt{PAFI\_}$V_{1,2,3}$ & 0 & 11 & 11 & 11 & 8 & 0 & 41 \\  
    \texttt{heated\_cell} %(4295, 4528, 4225, 4414, 5337 K)
    &   &   &   &   &   & 9 & 9 \\ 
    \hline 
    Total DB$_2$ & 1075 & 61 & 61 & 61 & 58 & 9 & 1325 \\
    \hline 
    \hline 
    {DB$_3$} & \multicolumn{6}{c}{} &  {} \\
    & & & & & & \\
    \hline 
    DB$_2$ & 1075 & 61 & 61 & 61 & 58 & 9 & 1325 \\
    \texttt{heated\_cell}   & & & & & & 388 & 388 \\
    \hline 
    Total DB$_3$ & 1075 & 61 & 61 & 61 & 58 & 397 & 1722 \\
    \hline 
    \hline 
  \end{tabular}}
\caption{An iterative list of atomic configurations for the minimal databases DB$_{1,2,3}$. \texttt{Cxx} denotes the deformations used to obtain accurate elastic constants. The cubic cell of the bcc lattice is subjected to various non-zero strains, $\epsilon$, for the class \texttt{$\epsilon$\_bulk}. The \texttt{noised\_} configurations, designed to mimic the MD of bulk, mono- and di-vacancies in various configurations are denoted by $V_1$ and $V_2$. \texttt{NEB} and \texttt{PAFI\_} represents sampling from 0 K to finite temperature for the vacancy jump, employing the NEB~\cite{neb,drag} and \textsf{PAFI} methods~\cite{SwinburneMarinica2018, pafi},respectively. 
\texttt{MDr} denotes the molecular dynamics trajectories for heating the system from 300 K to 5000 K, which provide the class \texttt{heated\_cell}. Further details about all the classes can be found in the text.
}
  \label{tab:dataset_all}
\end{table}

The \texttt{Cxx} class contains configurations involving iso-volumic deformations, from which the values of the bulk modulus $B$ and the anisotropic elastic constants $C_{11}$, $C_{12}$, and $C_{44}$ can be easily extracted. This class provides reliable information for the bcc elastic constants of the MLP. We have used 39 deformations. To minimize numerical round-off errors, the \textit{ab initio} energy calculations are performed in $(4 a_0)^3$ cubic supercells (128 atoms). 
The \texttt{$\epsilon$\_bulk} class corresponds to random deformation at a constant volume of the cubic cell of 2 atoms of bcc.  We impose a deformation $\boldsymbol{\epsilon}_{0}$ to which we add a random tensor $\delta \epsilon$ defined by $\delta \epsilon_{ij} {\sim} \varepsilon \: \mathcal{N}(0,1)$. $\varepsilon$ is the amplitude of random noise and $\mathcal{N}(0,1)$ is a standardized Gaussian distribution. In the end, we apply the following deformation tensor to the configuration :
$
	\boldsymbol{\epsilon} = \boldsymbol{\epsilon}_0 + \frac{1}{2} \left( \boldsymbol{\epsilon} + \boldsymbol{\epsilon}^{\top} \right),
$
We apply uniformly distributed deformations between $-5 \%$ and $5 \%$ with a random parameter $\varepsilon = 0.01$. In the end, we generate 1000 random deformed configurations.

The \texttt{noised\_} classes are designed to mimic molecular dynamics simulations at a given temperature and avoid the computational expense of \textit{ab initio} molecular dynamics. This is achieved by adding carefully thermal noise to the relaxed 0 K configurations of bulk, mono-, di-, and tri-vacancies.

The class \texttt{NEB\_} corresponds to standard Nudged Elastic Band~\cite{neb} pathways computed in DFT for the first nearest-neighbor migration of mono-, di- and tri-vacancies. The convergence criterion is defined as the maximum force being less than $10^{-2}$ eV/\AA\ . 
Once the first version of the potentials was fitted from DB$_1$, the MLP potentials were used to generate finite-temperature pathways from the 0 K trajectories. These configurations are included in the \texttt{PAFI\_} class. 
The finite temperature configurations are sampled from the \textsf{PAFI}~\cite{SwinburneMarinica2018} hyperplanes near the saddle point at a given temperature. 

The class \texttt{heated\_cell} corresponds to NPT molecular dynamics simulations at zero pressure, conducted from 300 K to 5000 K for a simulation cell containing perfect bulk bcc and A15, mono-, di-, and tri-vacancies. The heating ramp is applied at a rate of 5 K/ps. 
From the molecular dynamics performed with the MLP derived from DB$_1$, configurations were selected between 3000 K and 5000 K (if the potential was stable, see main text discussion). For DB$_3$, we randomly chose 388 configurations distributed between 1000 K and 4000 K to help stabilize the bcc-to-liquid as well as A15-to-liquid transitions.

\end{document}